\documentclass[a4paper,fleqn,usenatbib]{mnras}

\usepackage{graphicx}	
\usepackage{amsmath}	
\usepackage{amssymb}	
\usepackage{url}
\usepackage{color,soul}
\usepackage{multirow}
\usepackage{float}

\newcommand{\core}{\textit{COrE}}
\newcommand{\planck}{\textit{Planck}}

\title[Impact of modelling foreground uncertainties]{Impact of modelling foreground uncertainties on future CMB polarization satellite experiments}

\author[Herv\'ias-Caimapo et al.]{\parbox[t]{\textwidth}{Carlos Herv\'ias-Caimapo$^1$\thanks{E-mail: carlos.herviascaimapo@postgrad.manchester.ac.uk}, Anna Bonaldi$^{1,2}$ and Michael L. Brown$^1$} \vspace*{8pt} \\ 
$^1$Jodrell Bank Centre for Astrophysics, School of Physics \& Astronomy, University of Manchester, Oxford Road, Manchester M13 9PL, U.K. \\
$^2$SKA Organisation, Lower Withington Macclesfield, Cheshire SK11 9DL, U.K.}

\date{Accepted XXX. Received YYY; in original form ZZZ}

\pubyear{2017}

\begin{document}
\label{firstpage}
\pagerange{\pageref{firstpage}--\pageref{lastpage}}
\maketitle

\begin{abstract}
We present an analysis of errors on the tensor-to-scalar ratio due to residual diffuse foregrounds. We use simulated observations of a CMB polarization satellite, the Cosmic Origins Explorer, using the specifications of the version proposed to ESA in 2010 (\core{}). We construct a full pipeline from microwave sky maps to $r$ likelihood, using two models of diffuse Galactic foregrounds with different complexity, and assuming component separation with varying degrees of accuracy. Our pipeline uses a linear mixture (Generalized Least Squares) solution for component separation, and a hybrid approach for power spectrum estimation, with a Quadratic Maximum Likelihood estimator at low $\ell$s and a pseudo-$C_{\ell}$ deconvolution at high $\ell$s. In the likelihood for $r$, we explore modelling foreground residuals as nuisance parameters. Our analysis aims at measuring the bias introduced in $r$ by mismodelling the foregrounds, and to determine what error is tolerable while still successfully detecting $r$. We find that $r=0.01$ can be measured successfully even for a complex sky model and in the presence of foreground parameters error. However, the detection of $r=0.001$ is a lot more challenging, as inaccurate modelling of the foreground spectral properties may result in a biased measurement of $r$. Once biases are eliminated, the total error on $r$ allows setting an upper limit rather than a detection, unless the uncertainties on the foreground spectral indices are very small, i.e. equal or better than 0.5\% error for both dust and synchrotron. This emphasizes the need for pursuing research on component separation and foreground characterization  in view of next-generation CMB polarization experiments.
\end{abstract}

\begin{keywords}
cosmic background radiation -- inflation -- diffuse radiation -- early Universe
\end{keywords}


\section{Introduction} The successful detection of primordial Cosmic Microwave Background (CMB) polarization $B$-modes would confirm the inflationary paradigm, by probing the existence of gravitational waves that sets up the primordial tensor perturbations in the new born Universe during inflation. Also, this would allow us access into the energy scale of the very early Universe, $\sim10^{16}$ GeV. See \citet{kamionkowski_2016,eco_paper_inflation} for further details.

The astronomical community has put significant effort on the measurement of $B$-modes. 
Planned future satellite and balloon experiments, such as CORE \citep{core_paper}, LiteBIRD \citep{litebird_2014}, PIXIE \citep{pixie_2011}, PRISM \citep{prism_2014}, LSPE \citep{lspe_2012}, and ground-based experiments, such as SPT \citep{spt_2015}, BICEP2-Keck \citep{bicep2keck_2014}, POLARBEAR \citep{polarbear_2014}, among others, aim at detecting the large-scale $B$-mode polarization from the CMB in the near future. To accomplish this, the development of new detector technologies will allow an unprecedented high polarization sensitivity at microwave frequencies, capable of detecting $r \sim 10^{-3}$, if indeed the final error is dominated by instrument noise.

It is worth pointing out that the signal could be much smaller, which would definitively test the limits of our instrumentation and abilities.  Even if  this is not the case, however, achieving the required sensitivity is not enough, because of the presence of bright diffuse Galactic and extra-galactic foregrounds that block our clean view into the CMB. Therefore, component separation techniques are developed to model and subtract these foregrounds, in order to obtain the cleanest possible CMB maps. The question is then, how accurate can we model and clean the foregrounds to the level required for measuring $r=10^{-2}$--$10^{-3}$?

Several forecasts of tensor-to-scalar ratio measurements including foreground residuals have been performed for different experiments \citep{betoule_2009,armitage-caplan_2012,errard_2012,bonaldi_2014,remazeilles_2015,alonso_2016}. In this work, we study how the error in the diffuse foregrounds component separation modelling propagates into the tensor-to-scalar ratio. Our approach is quite agnostic from the point of view of physical modelling of the Galactic emission, and it focuses on quantifying the bias on $r$ corresponding to some arbitrary modelling error levels.  We also consider component separation and error mitigation techniques of different level of complexity.


This paper is organized as follows: In Section \ref{sec:simulations}, we introduce the model we use to create simulated observations of the microwave sky by a representative future CMB satellite. In Section \ref{sec:methods}, we describe the pipeline we use to forecast the bias on the tensor-to-scalar ratio. In Section \ref{sec:results}, we show the resulting $r$ bias for two different sky models, under different assumptions on component separation modelling complexity. Finally, in Section \ref{sec:conclusions}, we draw our conclusions.
\begin{figure*}
	\begin{center}
    	\includegraphics[width=0.5\textwidth]{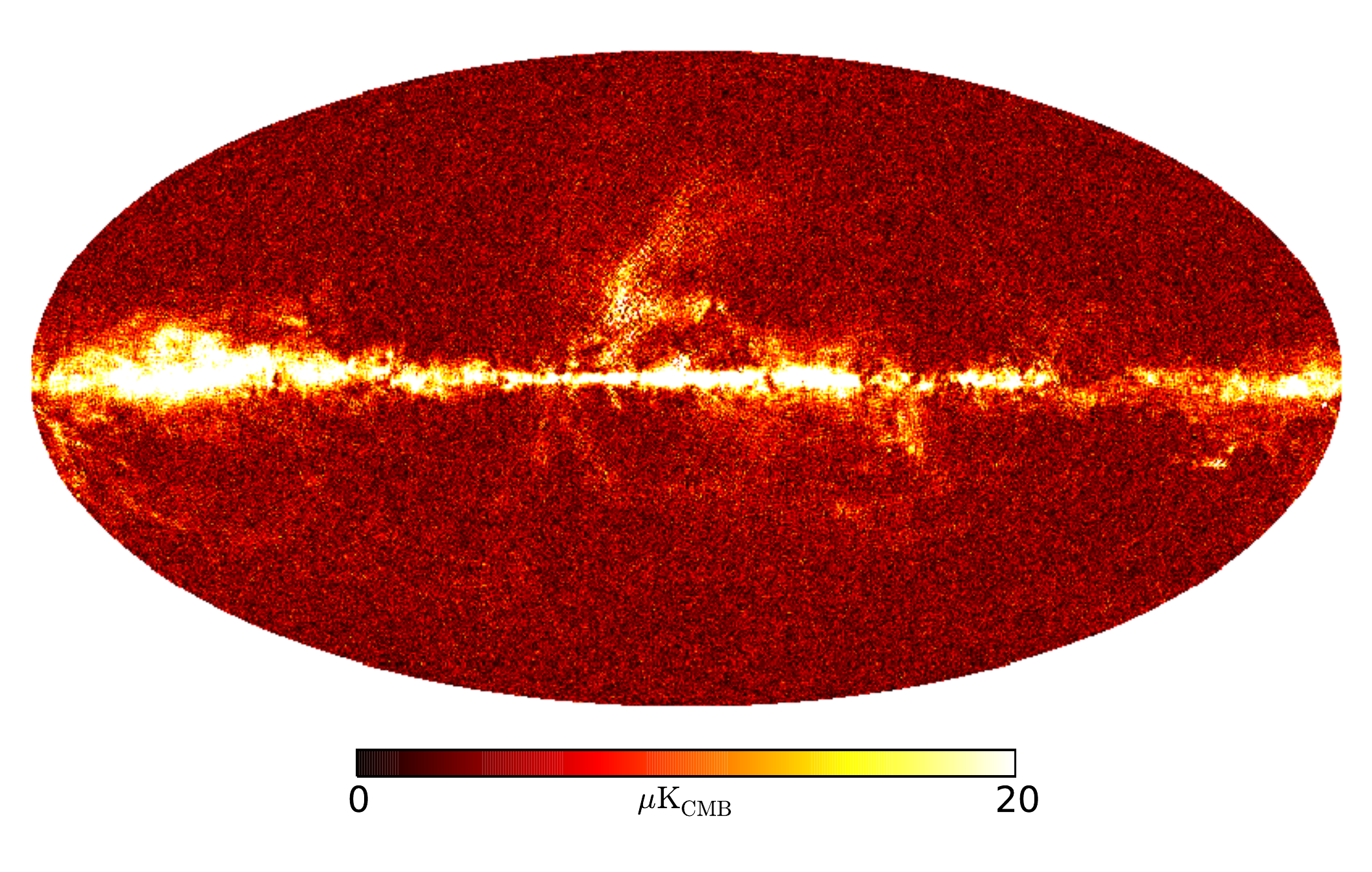}~
    	\includegraphics[width=0.5\textwidth]{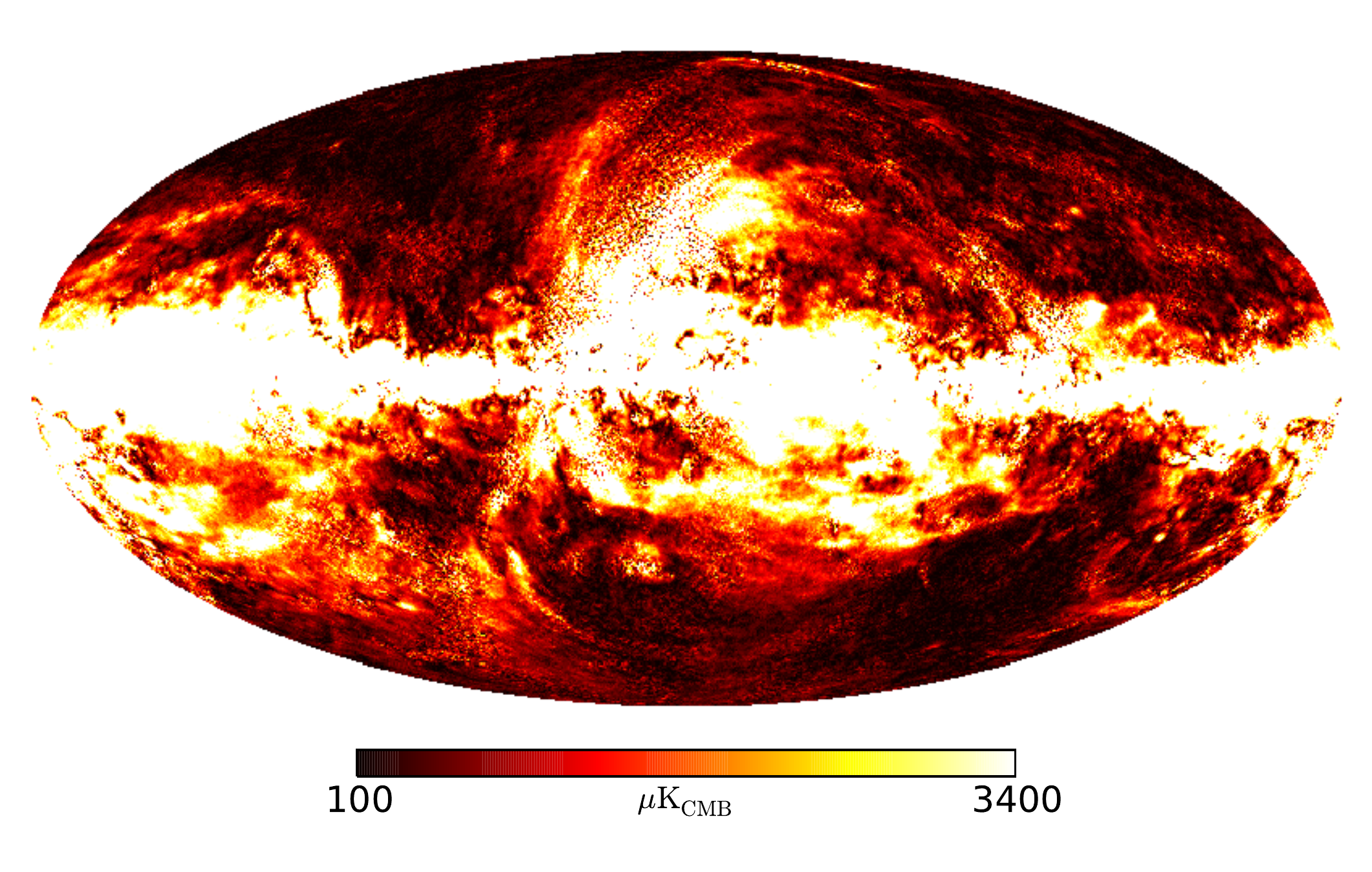}~
	\end{center}
    \caption{Polarization intensity $P=\sqrt{Q^2+U^2}$ maps of the simulated sky at 105\,GHz (left) and 555\,GHz (right), for the sky model with variable spectral indices. Both maps are  dominated by thermal dust emission. The maps of the model with spatially constant spectral indices look very similar. \label{fig:simulated_maps}}
\end{figure*}

\section{Simulated observations} \label{sec:simulations} For our analysis, we use the specifications of the Cosmic Origins Explorer (\core{}), as outlined in an earlier version of the experiment presented to ESA in 2010 \citep{core_paper}.

Recently the mission, renamed CORE, was formally submitted to ESA (CORE collaboration et al. in prep.) with different specifications (in particular more frequency channels, more tightly packed in the 60-600 GHz frequency range). However, in this work, we use the specifications from the earlier proposal to be representative of the capabilities of a future CMB polarization experiment. The frequencies, beam sizes, and sensitivities used in this work are listed in Table~\ref{table:core_specs}.

We perform our simulations using \textsc{healpix} \citep{gorski_2005} maps with a resolution parameter of $N_{\rm side}=512$, corresponding to a pixel size of $\sim 7$\,arcmin. Some of the actual \core{} bands have better resolution than the one allowed by such a pixel size, so we limit the band resolution to 7\,arcmin in these frequencies, marked with $^*$ in Table~\ref{table:core_specs}. This modification does not change our results appreciably because we focus on diffuse foreground components and primordial $B$-modes, both dominant at low multipoles.

To simulate the full-sky observations of the microwave sky, we use the model presented in \citet{hervias_2016}, based on the polarization results from the 2015 data release of \planck{} \citep{planck_2015_10}. 

\begin{table*}
	\centering
    \begin{tabular}{|c|c|c|c|c|c|c|c|c|c|c|c|c|c|c|c|}
	    \hline
	    Band [GHz] & 45 & 75 & 105 & 135 & 165 & 195 & 225 & 255 & 285 & 315 & 375 & 435 & 555 & 675 & 795  \\
		\hline    
	    Beam FWHM [arcmin] & 23.3&14.0&10.0&7.8&$7.0^*$&$7.0^*$&$7.0^*$&$7.0^*$&$7.0^*$&$7.0^*$&$7.0^*$&$7.0^*$&$7.0^*$&$7.0^*$&$7.0^*$ \\
	    \hline
	    Noise [$\mu {\rm K}_{\rm A}\cdot$arcmin]& 8.61&4.09&3.5&2.9&2.38&1.84&1.42&2.43 & 2.94 & 5.62 & 7.01 & 7.12 & 3.39 & 3.52 & 3.60 \\
	    \hline
    \end{tabular}
    \caption{\core{} satellite specifications used in this work to simulate observations, taken from \citet{core_paper}. As explained in the main text, the bands marked with a $^*$ have better resolution than 7\,arcmin, but have been simulated with a 7\,arcmin pixel size ($N_{\rm side}=512$ Healpix maps) to limit the computational complexity of our analysis. \label{table:core_specs}}
\end{table*}
In this work, we consider three polarized sky components: CMB, thermal dust, and synchrotron.  The main features of the sky model are as follows:

\begin{itemize}
\item \textbf{CMB} It is a Gaussian realization of a theory power spectrum produced with \textsc{CAMB} \citep{howlett_2012}. The adopted cosmology is the following: $T_{\rm CMB}=2.725$\,K, $\Omega_b=0.0461$, $\Omega_{c}=0.2286$, $\Omega_{\Lambda}=0.724$, $\Omega_{\nu}=0.0013$, $H_0=70$\,km/s/Mpc, $\tau=0$, $n_s=0.96$ and $n_t=0$. We include tensor perturbations with two different values of the tensor-to-scalar ratio $r=0.01$ and $r=0.001$. Our simulation includes lensing $B$-modes, generated from the \textsc{CAMB} power spectrum.

\item \textbf{Thermal dust} We use the dust polarization template described in \citet{hervias_2016}, smoothed to $1^{\circ}$. As a spectral law, we use a modified black body with a constant temperature of $T_d=21$\,K. For the $\beta_{\rm dust}$ spectral index, we use two models: one constant ($\beta_{\rm dust}=1.53$) and one spatially variable (based on the thermal dust spectral index map presented in \citet{planck_2015_10} and smoothed to $3^{\circ}$).

\item \textbf{Synchrotron} We use the syncrotron polarization template described in \citet{hervias_2016}, smoothed to $1^{\circ}$. We use a power law frequency scaling with a $\beta_{\rm syn}$ spectral index. Again we use either a constant ($\beta_{\rm syn}=3.1$, as used in \citealt{planck_2015_10}), or a spatially variable (\citealt{giardino_2002}, having a resolution of $10^{\circ}$) spectral index.
\end{itemize}
Some maps for the sky model with spatially-variable indices are shown in 
Fig.~\ref{fig:simulated_maps}. 

For each model, we produce 100 sets of \textsc{fits} maps of the observed sky at each band.  Each set has the frequency bands, resolution and white noise levels as specified in Table~\ref{table:core_specs}. Each set has a different CMB and white noise realization, but the same foreground components. We produce them with a \textsc{healpix} resolution parameter of $N_{\rm side}=512$. We also produce 100 low-resolution sets with $N_{\rm side}=16$. In this case, the modelled sky is produced with a resolution of $3.5^{\circ}$ across all bands, according with the larger size of the pixels. Although the beam size is not very well sampled by this pixel size, we have verified that, once both beam and pixel window function are deconvolved, the pipeline described in Sec. \ref{sec:powerspectra} yields an unbiased recovery of the CMB polarization power spectra.
Since the seed used to create the CMB realization for a given set is the same always, the $N_{\rm side}=512$ and $N_{\rm side}=16$ CMB maps are the same realization, but with different resolution.

\section{Methodology} \label{sec:methods} In this section, we describe the various steps of our pipeline: component separation, power spectrum and likelihood estimation. 

\subsection{Component separation} To perform the component separation, we  rely on the linear mixture model, stated as follows. 
The intensity of each foreground $j$ at a frequency band $\nu$ and in a line of sight $p$ can be expressed as $a_j(\nu) s_j(p)$, where $a_j(\nu)$ is the corresponding assumed spectral law and $s_j(p)$ would correspond to the template map of each foreground at a fixed arbitrary frequency. Then, the observed intensity $y(\nu,p)$ is 
\begin{equation}
	y(\nu,p) = (\sum_{j} a_j(\nu) s_j(p))*B(\nu) + n(\nu,p) \text{,}
\end{equation}
where $B(\nu)$ is the instrumental beam depending on the frequency channel $\nu$, $*$ denotes convolution and $n(\nu,p)$ is the instrumental noise. If the resolution of all frequency channels is the same, for each line of sight it is possible to rewrite the previous equation in matrix notation,
\begin{equation}
	\bmath{y} = \bmath{\sf A} \bmath{s} + \bmath{n} \text{,}
\end{equation}
where $\bmath{\sf A}$ is the mixing matrix, with dimensions $N_c$ (number of components) times $N_b$ (number of spectral bands). The vector $\bmath{s}$ now contains all the components $s_j$ convolved by the frequency-constant beam $B$ and $\bmath{y}$ contains all the data maps $y$.

If the linear mixture models holds, it is possible to obtain an estimate of the components with a suitable linear mixture of the frequency channels, $\bmath{s}=\bmath{\sf W} \bmath{y}$. If we know the mixing matrix, one possible solution is the Generalized Least Square solution (GLS), given by the matrix $\bmath{\sf W} = \left[ \bmath{\sf A}^{\dagger} \bmath{\sf C_n}^{-1} \bmath{\sf A} \right]^{-1} \bmath{\sf A}^{\dagger} \bmath{\sf C_n}^{-1}$, where $\bmath{\sf C_n}$ is the covariance matrix of the instrumental noise. This solution is unbiased in recovering $\bmath{s}$, but retains a noise contribution. However, it minimizes the variance of the error when the sky signal $\bmath{s}$ is deterministic \citep{delabrouille_2009}. In practice, an estimate of the mixing matrix $\bmath{\sf A}$ is typically calculated by parametrizing the spectral laws of the CMB and foreground components and by estimating the relevant parameters from the data. In this paper, we skip such estimation: we assume some error on the spectral parameters describing the true mixing matrix and propagate them through the full pipeline. 

As stated above, one important assumption of the linear mixture model (at least when applied in pixel domain) is that the instrumental beam does not depend on frequency. This is not true in general, nor it is  for \core{}, as shown by Table~\ref{table:core_specs}. To overcome this problem, we pre-processed all maps by smoothing them with a Gaussian beam, thus equalizing their resolution to  23.3\,arcmin (that of the lowest frequency channel, for the high-resolution sets $N_{\rm side}=512$) or $3.5^{\circ}$ (the resolution sampled by the $N_{\rm side}=16$ maps for the low-resolution sets).

\subsection{Power spectra estimation}\label{sec:powerspectra} We estimated the polarization power spectra from the CMB maps with a hybrid approach: using a Quadratic Maximum Likelihood (QML) estimator at low ($\ell < 30$) multipoles and a pseudo-$C_{\ell}$ estimator at the remaining intermediate and high multipoles. The QML estimator is optimal at low multipoles, and it is able to recover the reionization bump at $\ell < 10$.  However, it  gets very computationally demanding very quickly with increasing resolution. The pseudo-$C_{\ell}$ estimator is appropriate for high multipoles, which are unobtainable for the QML estimator, where it can recover the first acoustic peak at $\ell \sim 100$. This hybrid approach has been shown to be nearly optimal in the whole $\ell$ range and at the same time computationally feasible \citep[e.g.][]{Efstathiou2004,efstathiou_2006}. The simulated observations at $N_{\rm side}=512$ are used for estimating the pseudo-$C_{\ell}$ power spectra, while the low-resolution maps with $N_{\rm side}=16$ are used for the QML estimator.

The QML method we use is based on \citet{tegmark_1997,tegmark_2001}, see also \citet{Efstathiou_2004_2,Gruppuso2009}. It works on pixel space, constructing an estimator based on the covariance matrices of the data. This method gives minimal error bars but it is very computationally demanding, since it requires operations of order $\mathcal{O}(N_{\rm d}^3)$, where $N_{\rm d}$ is the number of pixels outside the mask.

The pseudo-$C_{\ell}$ deconvolution method we use is described in \citet{brown_2005} and \citet{brown_2009}, which extended to polarization the technique proposed by \citet{hivon2002}. This method uses a fast spherical harmonic transform to estimate the pseudo-$C_{\ell}$ spectra on the masked sky, and corrects them for the effect of the sky cut, noise and filtering with a deconvolution process. The output power spectrum is binned with 
bandpass window functions $W_{b\ell}$, and needs to be compared to a binned theory power spectrum
\begin{equation}
	\bmath{P}_{b} = \sum_{\ell} \frac{\ell(\ell+1)}{2\pi} \frac{W_{b\ell}}{\ell} \bmath{C}_{\ell} \text{.}
\end{equation}
in the likelihood for $r$.
\begin{figure}
	\centering
    \includegraphics[width=1.0\columnwidth]{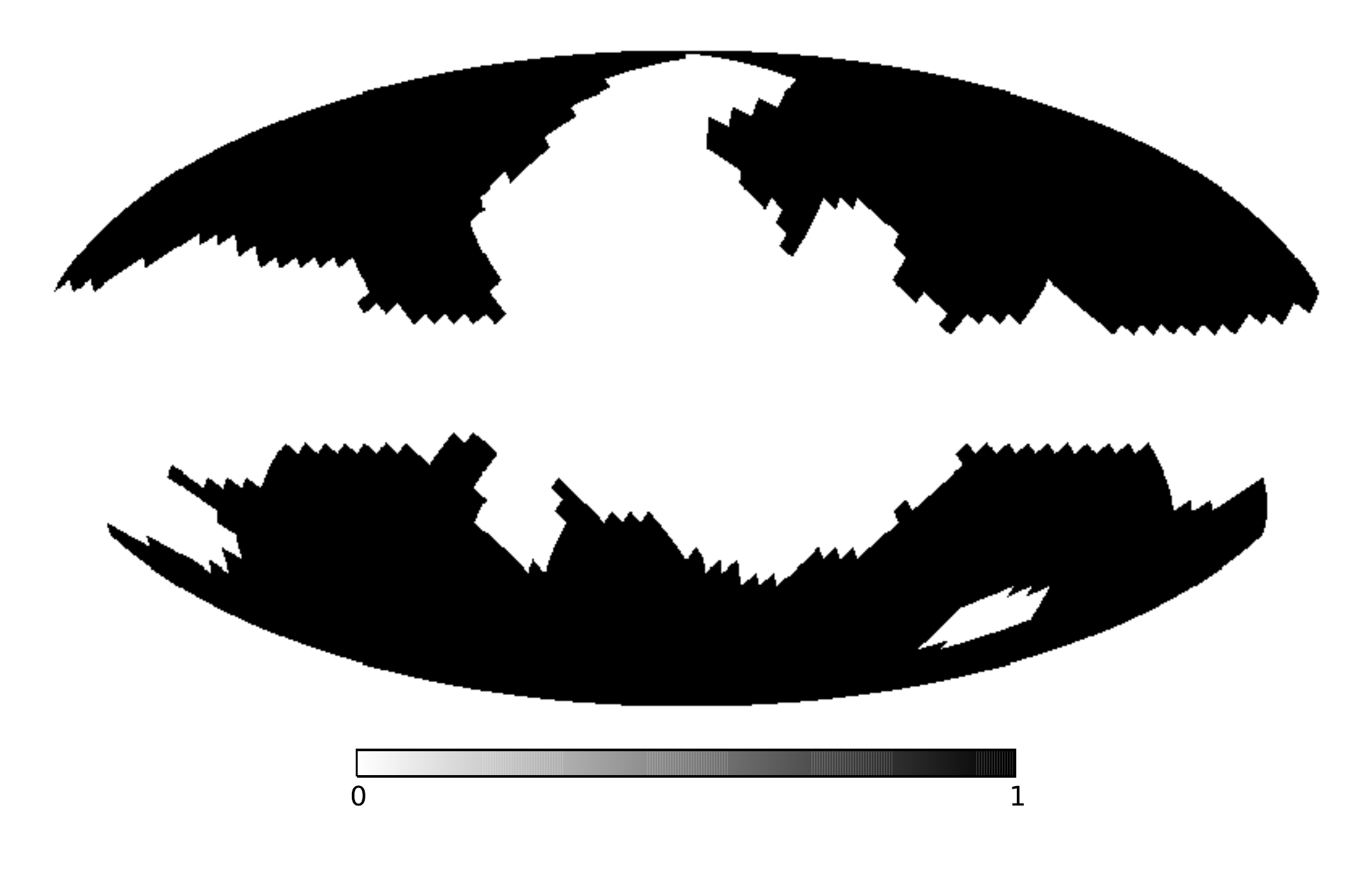}
    \includegraphics[width=1.0\columnwidth]{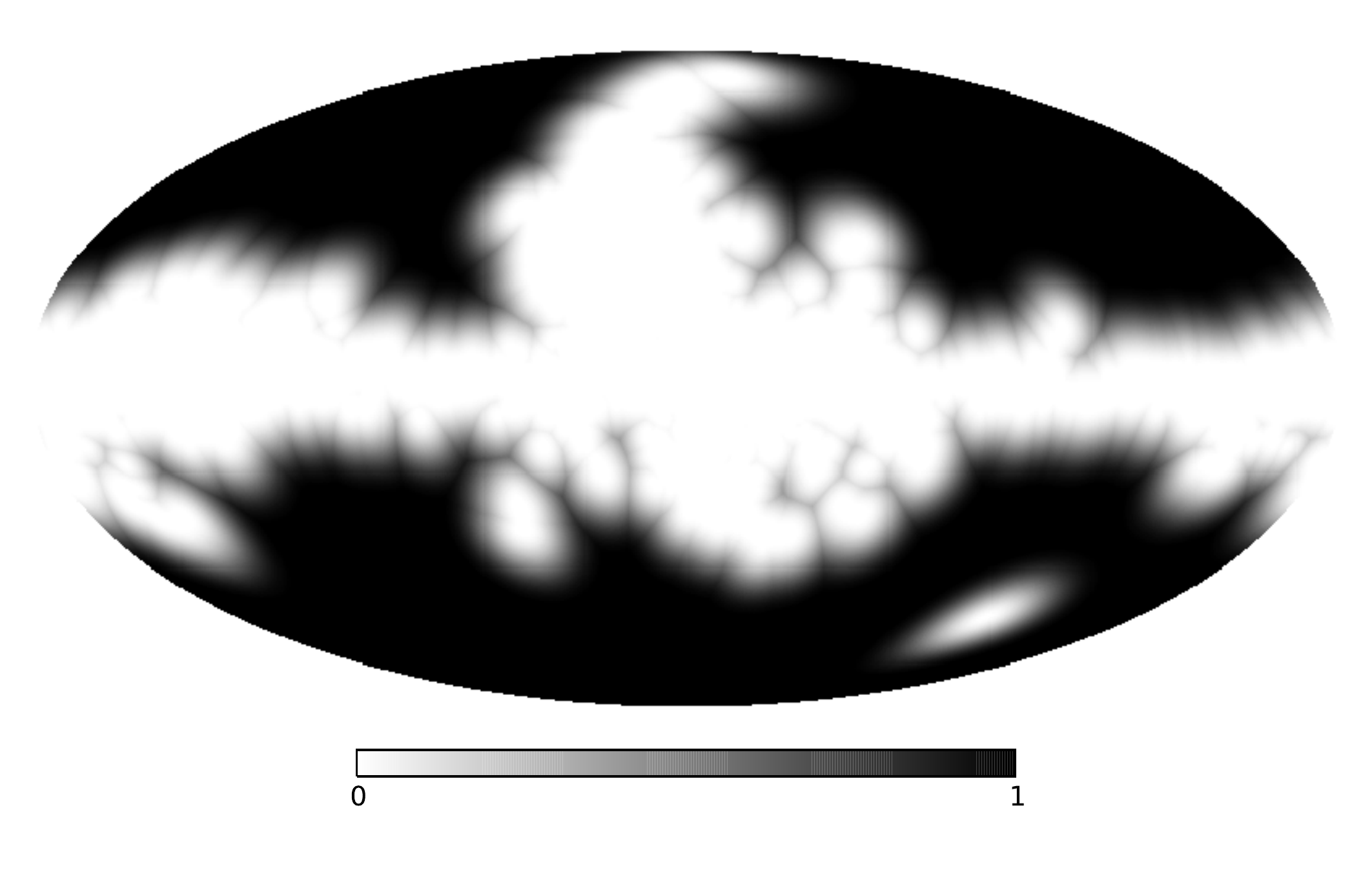}
    \caption{Default Galactic mask used for the power spectrum estimation, retaining a fraction of the sky $f_{\rm sky}=0.513$. Top: $N_{\rm side}=16$ mask used for the QML power spectrum estimation; bottom: $N_{\rm side}=512$ apodized mask used for the pseudo-$C_{\ell}$ power spectrum estimation. \label{fig:base_mask}}
\end{figure}

\subsubsection{Galactic mask} To exclude the foreground residual contamination due to the Galactic emission, we estimate the power spectrum outside a Galactic mask. The default mask is constructed using the dust and synchrotron polarization templates from \citet{planck_2015_10}, smoothing them to a FWHM of $3^{\circ}$, and masking every pixel with an intensity of $14 \mu$\,K or higher. We repeat this procedure for both $Q$ and $U$ maps, and dust and synchrotron. We combine all of them to produce the final mask. For the pseudo-$C_{\ell}$ power spectrum estimation, it is beneficial to use an apodized mask, because sharp edges make the deconvolution kernel more complicated.  Therefore, we apodize the  $N_{\rm side}=512$ mask by using the function 
\begin{equation}
	f(d) = \begin{cases} 1 - \cos^3(\frac{d\pi}{2s}) & d \le s \\ 1 & \text{otherwise} \end{cases}
\end{equation}
where $d$ is the distance between the pixel of interest and the closest masked pixel (with value 0), and $s$ is the distance scale of apodization (the scale in which the function goes from 1 to 0, $s=20^{\circ}$ in our case). The resulting apodized mask is shown in Fig.~\ref{fig:base_mask}, bottom. The sky fraction retained is $f_{\rm sky}=0.513$. The $N_{\rm side}=16$ version of this mask, needed for the QML estimator, is not apodized, and has been constructed by rounding the $N_{\rm side}=512$ mask and degrade to $N_{\rm side}=16$. This mask is shown in Fig.~\ref{fig:base_mask}, top.

\begin{figure*}
	\begin{center}
    	\includegraphics[width=0.5\textwidth]{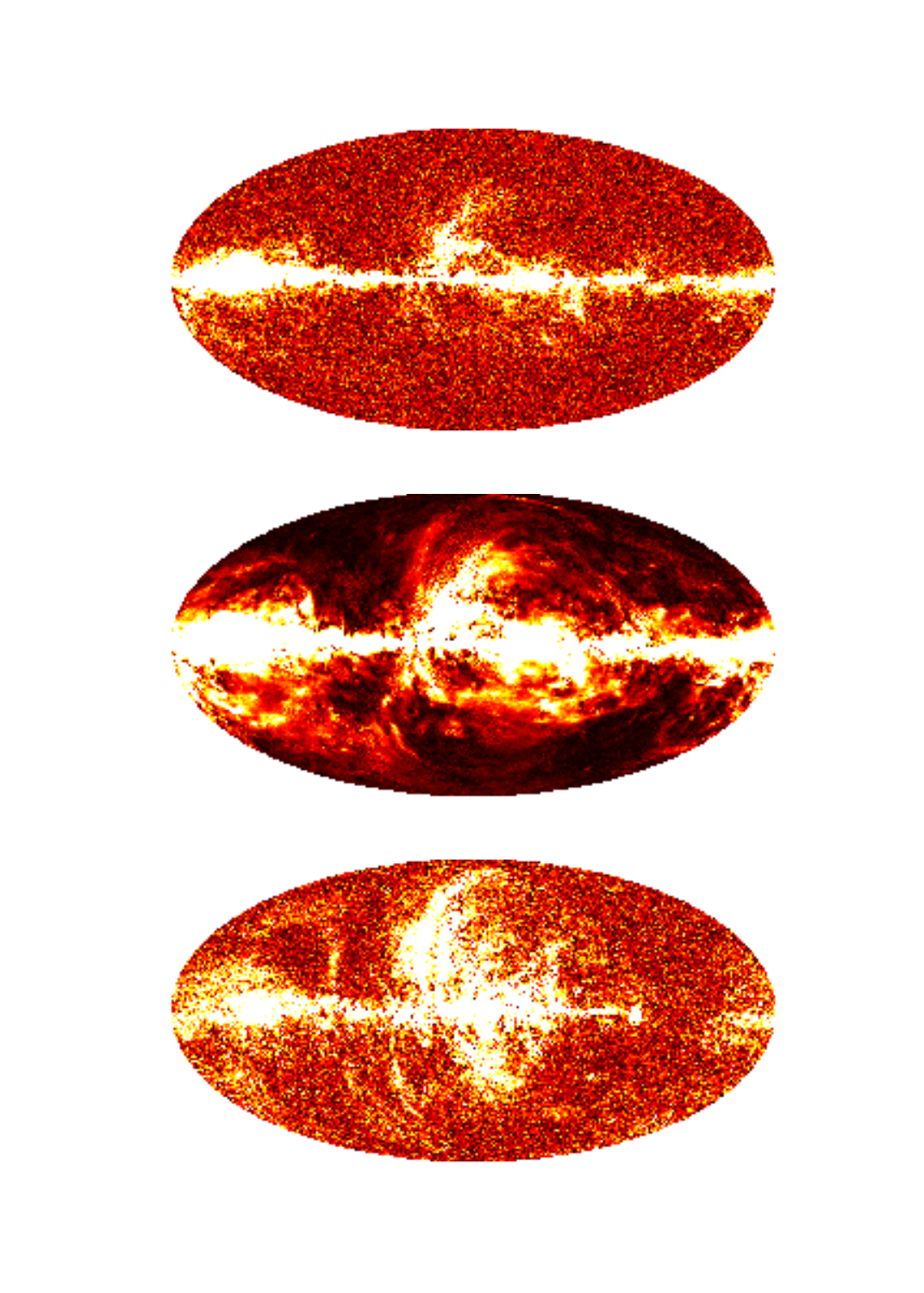}~
		\includegraphics[width=0.5\textwidth]{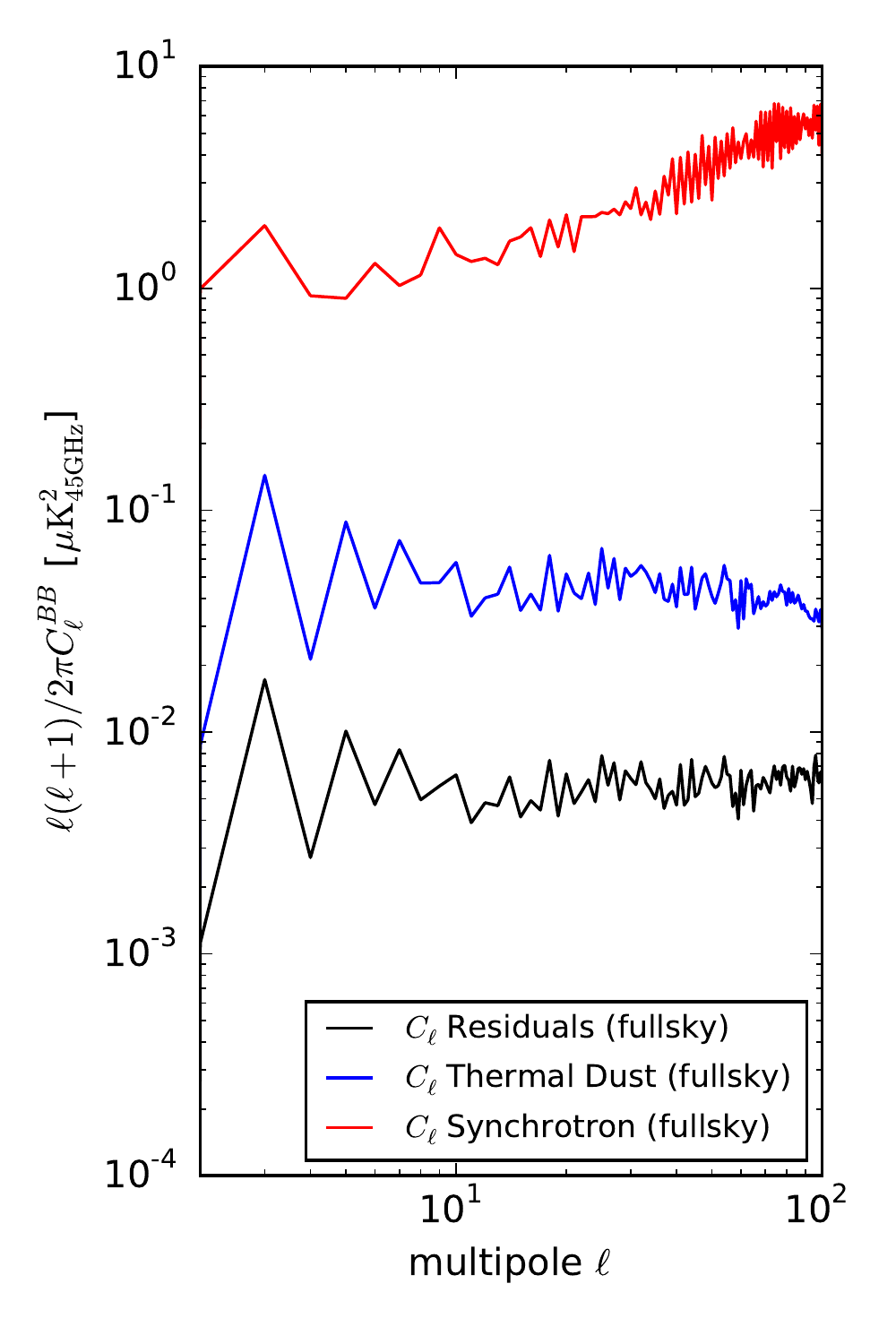}~
	\end{center}
    \caption{Left: polarization intensity maps of the foreground residuals (reconstructed-true CMB, top) compared to the thermal dust (middle) and synchrotron (bottom) maps reconstructed by the component separation. Right: full-sky $BB$ power spectrum of foregrounds residuals compared to the full-sky power spectrum of the reconstructed thermal dust and synchrotron foregrounds. Notice the similar shape between the residuals and the thermal dust.  \label{fig:modelling_foreground_residuals}}
\end{figure*}

\subsection{Cosmological parameters likelihood} We calculate the likelihood on the power spectra averaged over the 100 realizations of simulated observations, where we varied both the CMB and noise realizations. This effectively eliminates the cosmic variance bias, and only leaves the foreground residuals bias, which is of our interest. We define a standard Gaussian $\chi^2$ likelihood to calculate the posterior distribution of the tensor-to-scalar ratio. We define the $\chi^2$ as
\begin{equation}
	\chi^2(r) = \sum_{b b'} [P_b^{BB} - C_{b}^{BB, \rm theory}(r)] \bmath{\sf C}_{bb'}^{-1} [P_{b'}^{BB} - C_{b'}^{BB, \rm theory}(r)] \text{,} \label{eq:chi2}
\end{equation}
where $P_b^{BB}$ is the measured $B$-mode bandpower at bin $b$, $C_{b}^{BB, \rm theory}(r)$ is the binned $B$-mode theory spectrum and $\bmath{\sf C}_{bb'}^{-1}$ is the inverse of the binned signal+noise covariance matrix.
We construct $P_b^{BB}$ from the low-multipole and high-multipole analysis, by joining at $\ell=30$ (with no overlap) the results from the QML and the pseudo-$C_{\ell}$ estimators.


The theory power spectrum is binned using the bandpass window functions at high multiple range and using a top hat function centered at each bin in the low multipole range. The theory power spectrum is calculated as 
\begin{equation}
	C_{\ell}^{BB, \rm theory}(r) = \frac{r}{r_{\star}} C_{\ell}^{BB, \rm prim}(r_{\star}) + C_{\ell}^{BB, \rm lensing} \text{,} \label{eq:theory_cl}
\end{equation}
where $C_{\ell}^{BB,\rm prim}$ is the primordial (scalar+tensor perturbations) power spectrum at a given $r$, and $C_{\ell}^{BB,\rm lens}$ is the weak gravitational lensed power spectrum, which we assume as known.

The covariance matrix is calculated using the 100 realizations signal+noise complete runs of the pipeline (including the component separation). Therefore, it accounts for cosmic and noise variance but also foreground residuals effects.

\subsubsection{Modelling foreground residuals with nuisance parameters} 
The likelihood presented in equation~(\ref{eq:chi2}) assumes that the measured power spectra contain only CMB and noise. In reality, there are also some foreground residuals, due to non-perfect component separation. We are now going to extend this likelihood to explicitly model a foreground residual contribution 
\begin{equation}
	C_b^{BB,\rm new}(r) = C_b^{BB,\rm theory}(r) + A_{\rm dust}C_b^{BB,\rm dust} + A_{\rm syn}C_b^{BB, \rm syn} \text{,} \label{eq:theory_cl_2}
\end{equation}
where $C_b^{BB,\rm dust}$ and $C_b^{BB, \rm syn}$ are models for the $BB$ power spectra of synchrotron and dust residuals, respectively, and the amplitudes $A_{\rm dust}$ and $A_{\rm syn}$ are two free nuisance parameters that can be varied, together with $r$, and finally marginalized over. The need for adding one or both such extra parameters can be checked by seeing whether they improve the fit, by means of the reduced $\chi^2$ value.

In practice, a way to derive the foreground residual template models $C_b^{BB,\rm dust}$ and $C_b^{BB, \rm syn}$ is to assume that they are proportional to the dust and synchrotron power spectra. In our case, these can be computed from the foreground maps which are output of the component separation, as exemplified by Fig.~\ref{fig:modelling_foreground_residuals}. 


In the analysis that follows, $C_b^{BB,\rm dust}$ and $C_b^{BB, \rm syn}$ are the binned power spectra of the thermal dust and synchrotron, respectively, reconstructed by the component separation. We process these maps through the same procedure we use for the reconstructed CMB, that is, the power spectra estimation with pseudo-$C_{\ell}$ for the high-resolution map and the QML estimator for the low-resolution map, under the same conditions. 



\begin{table*}
\centering
    \begin{tabular}{ l l l l l}
    \hline
    \multicolumn{2}{| c |}{\textbf{Simulation run}} & \textbf{Sky model}& \textbf{Component separation model} & \textbf{Reference}\\
    \hline
    \multirow{2}{*}{Simple model}	& $r=0.01$	& \multirow{2}{*}{Spatially constant $\beta_{\rm dust}$, $\beta_{\rm syn}$}	& \multirow{2}{*}{Constant $\beta$s, with a $\pm$1,2,3\% error} 	& \multirow{2}{*}{\ref{sec:constant_constant}; Table~\ref{table:constant}}\\
    								& $r=0.001$	& 	& \\
    \hline
    \multirow{3}{*}{Complex model}	& $r=0.01$ 	& \multirow{3}{*}{Spatially variable $\beta_{\rm dust}$, $\beta_{\rm syn}$}	& \multirow{2}{*}{Constant $\beta$s (the average of the true variable $\beta$ maps)} & \multirow{2}{*}{\ref{sec:variable_constant}; Table~\ref{table:results_variable}, top}\\
    								& $r=0.001$ &	& \\
     								& $r=0.001$ &	& Variable $\beta$s, with a global error of 1\% and 0.5\% & \ref{sec:variable_variable}, Table~\ref{table:results_variable}, bottom \\    
	\hline
	\end{tabular}
    \caption{Summary of the different runs performed in this work.\label{table:summary}}
\end{table*}

\section{Results} \label{sec:results} We run the component separation pipeline, described in Section~\ref{sec:methods}, for the two sky models, with constant and spatially-variable spectral indices. The summary of all the runs performed in this paper, together with descriptions and the referenced section where the results appear, is shown in Table~\ref{table:summary}.



\begin{figure*}
	\centering
    \includegraphics[width=1.0\textwidth]{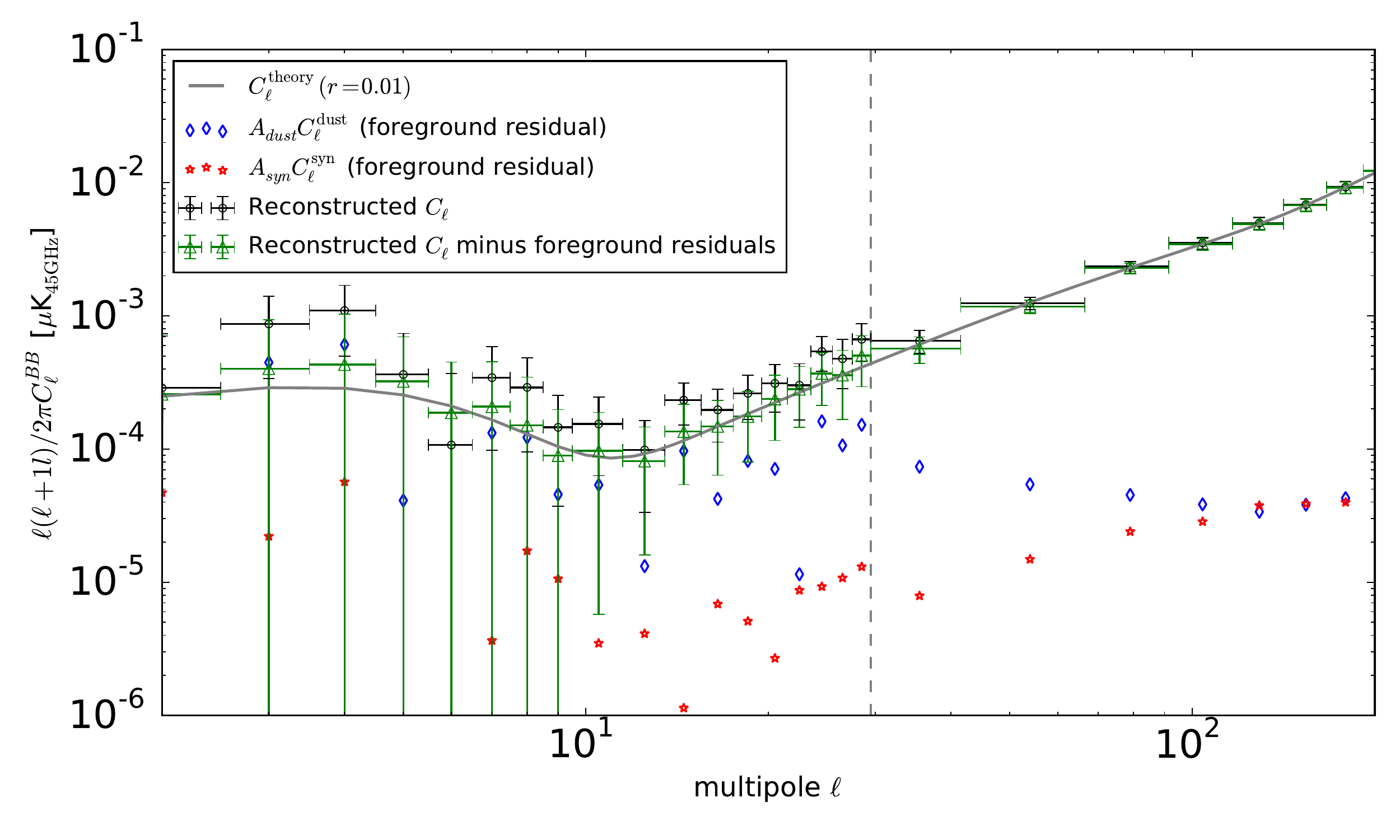}
    \caption{Reconstructed $BB$ power spectrum for the simulation with $r=0.01$, constant foreground spectral indices and a $+2$\% estimation error on both $\beta_{\rm dust}$ and $\beta_{\rm syn}$. The reconstructed CMB (on 10 frequency bands, with $\nu \leq 315$\,GHz) is biased (black circles). The modelled foreground residuals are shown as the diamonds and stars. 
The reconstructed CMB minus the modelled foregrounds residuals is shown as the green triangles, which can be compared with the theory power spectrum, shown as the grey curve. The multi-parameter likelihood yields $r=0.0088 \pm 0.0020$ \label{fig:ps_constant}}
\end{figure*}

\begin{figure*}
	\begin{center}
    	\includegraphics[width=0.5\textwidth]{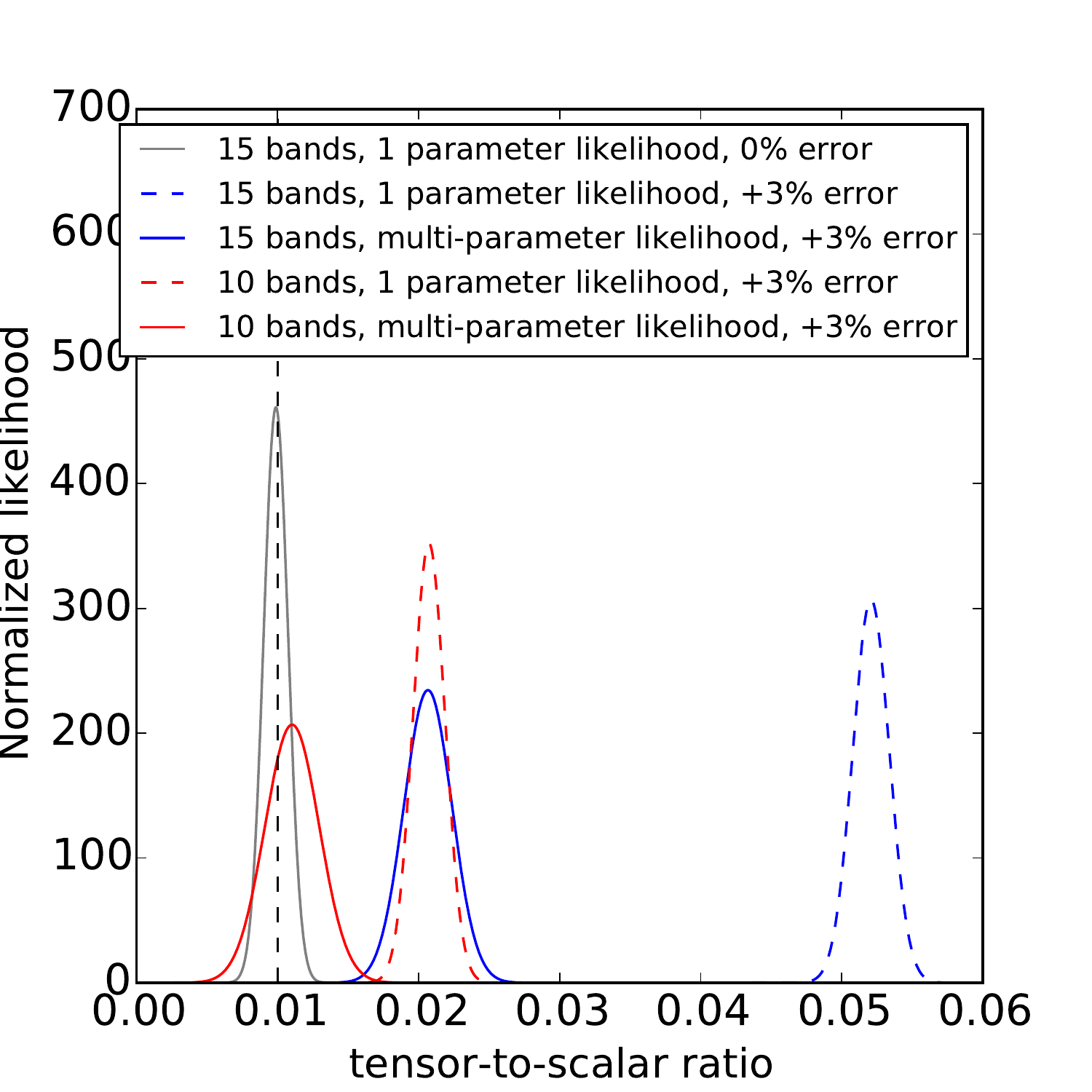}~
        \includegraphics[width=0.5\textwidth]{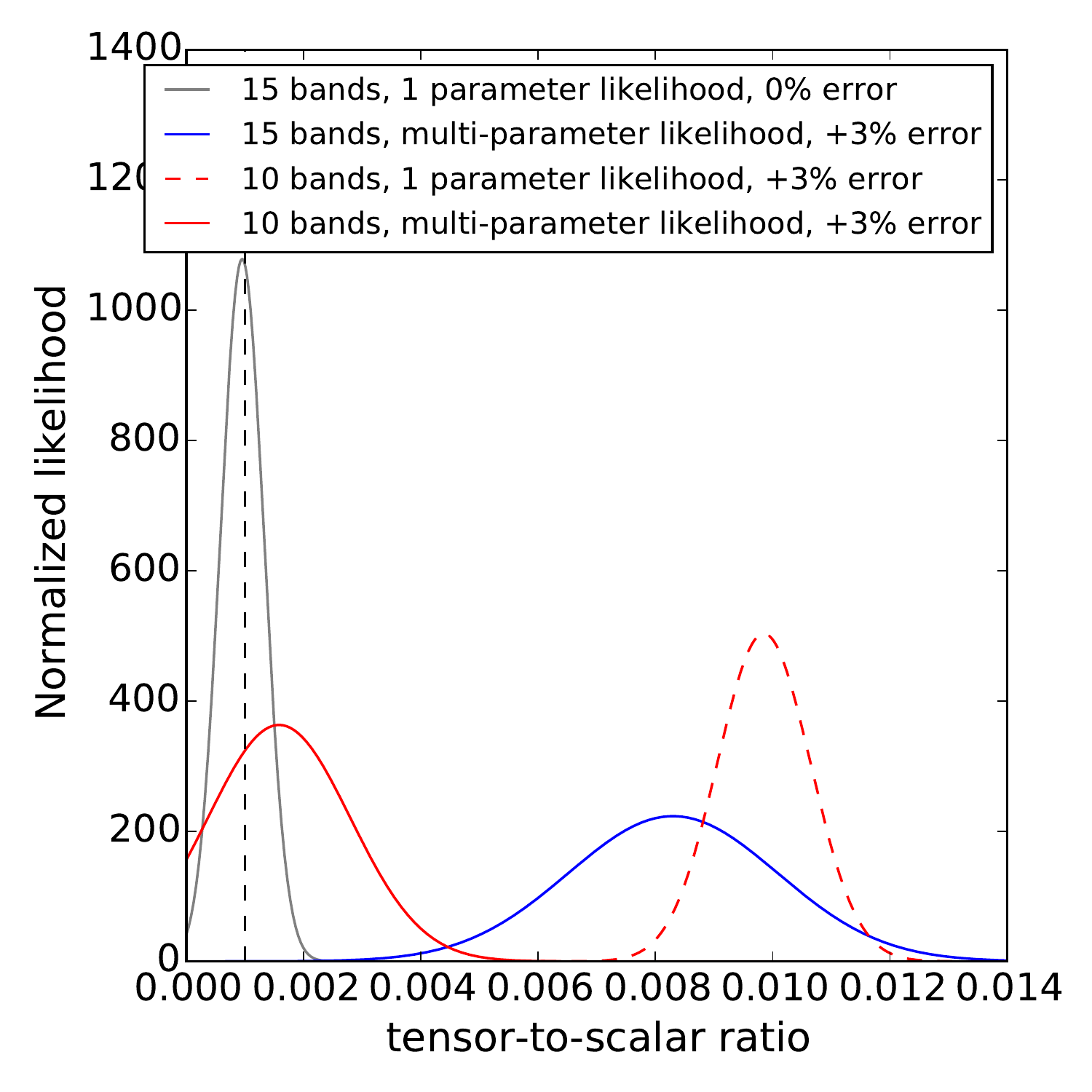}~
	\end{center}
	\caption{Left: tensor-to-scalar ratio likelihoods for the model with constant spectral indices and $r=0.01$. The dashed grey curve shows the likelihood for the perfect knowledge of foreground spectral indices, centered in the correct $r$. The dashed curves show the likelihood when an error of $+3$\% is made on both spectral indices, using the entire frequency range (blue) and limited ($\nu \leq 315$\,GHz) one (red). The solid curves show the results in the same cases when the multi-parameter likelihood is used. Right: same as the left panel, for the model with constant spectral indices and $r=0.001$. We do not show the 15 bands, 1 parameter likelihood case with $+3$\% error, since it is extremely biased, measuring $r=0.0470 \pm 0.0013$. \label{fig:likelihoods_constant}}
\end{figure*}

\begin{figure}
	\centering
	\includegraphics[width=1\columnwidth]{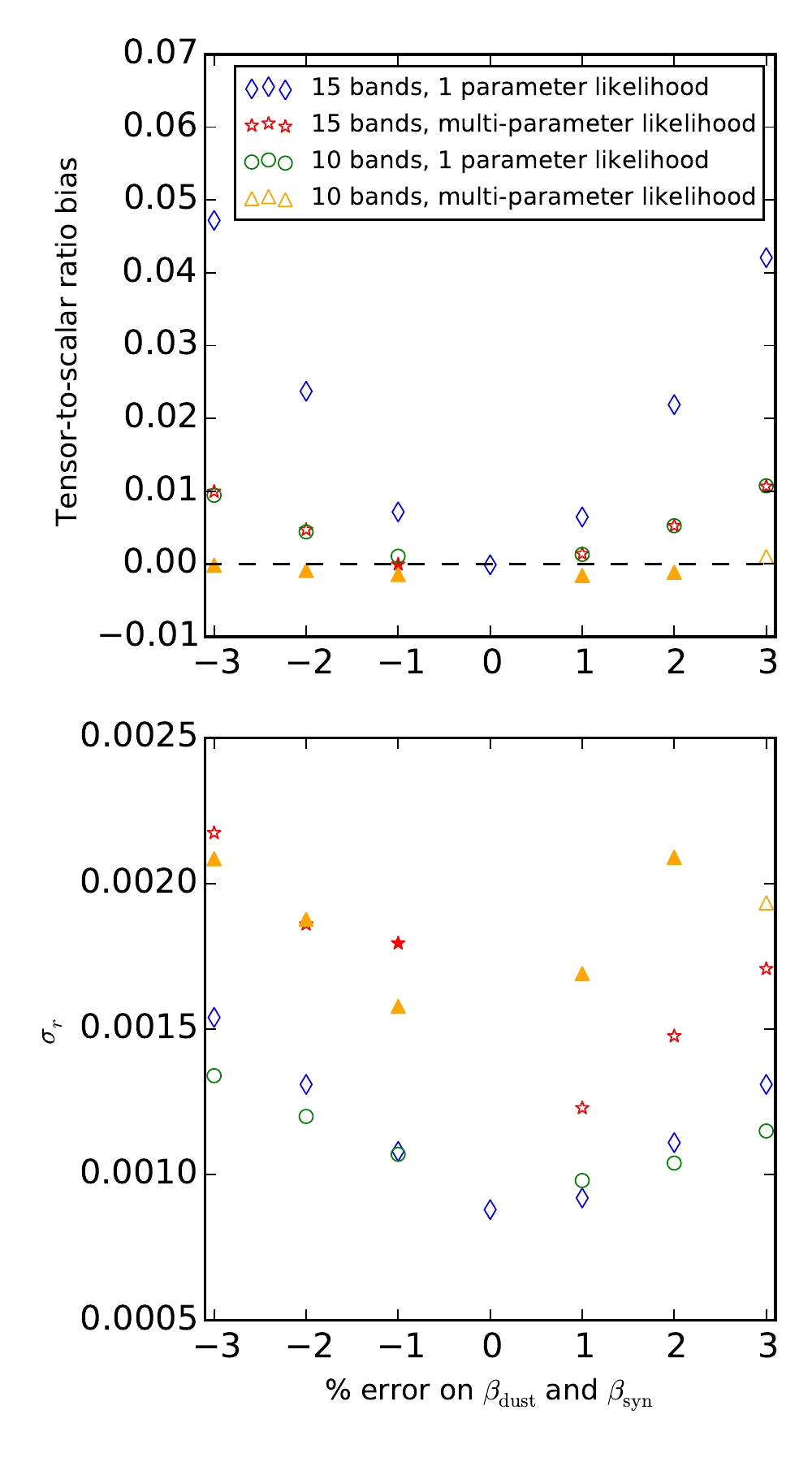}
	\caption{Tensor-to-scalar ratio bias (estimated -- true $r$, top) and error ($\sigma_r$, bottom) for different cases of fixed constant errors on both spectral indices, for the simulation with constant spectral indices and $r=0.01$. In the case of multi-parameter likelihoods, the empty symbols are the ones using only $A_{\rm dust}$, and the filled symbols are the ones using both $A_{\rm dust}$ and $A_{\rm syn}$. \label{fig:bias_r2_constant}}
\end{figure}

\begin{figure}
	\centering
    \includegraphics[width=1\columnwidth]{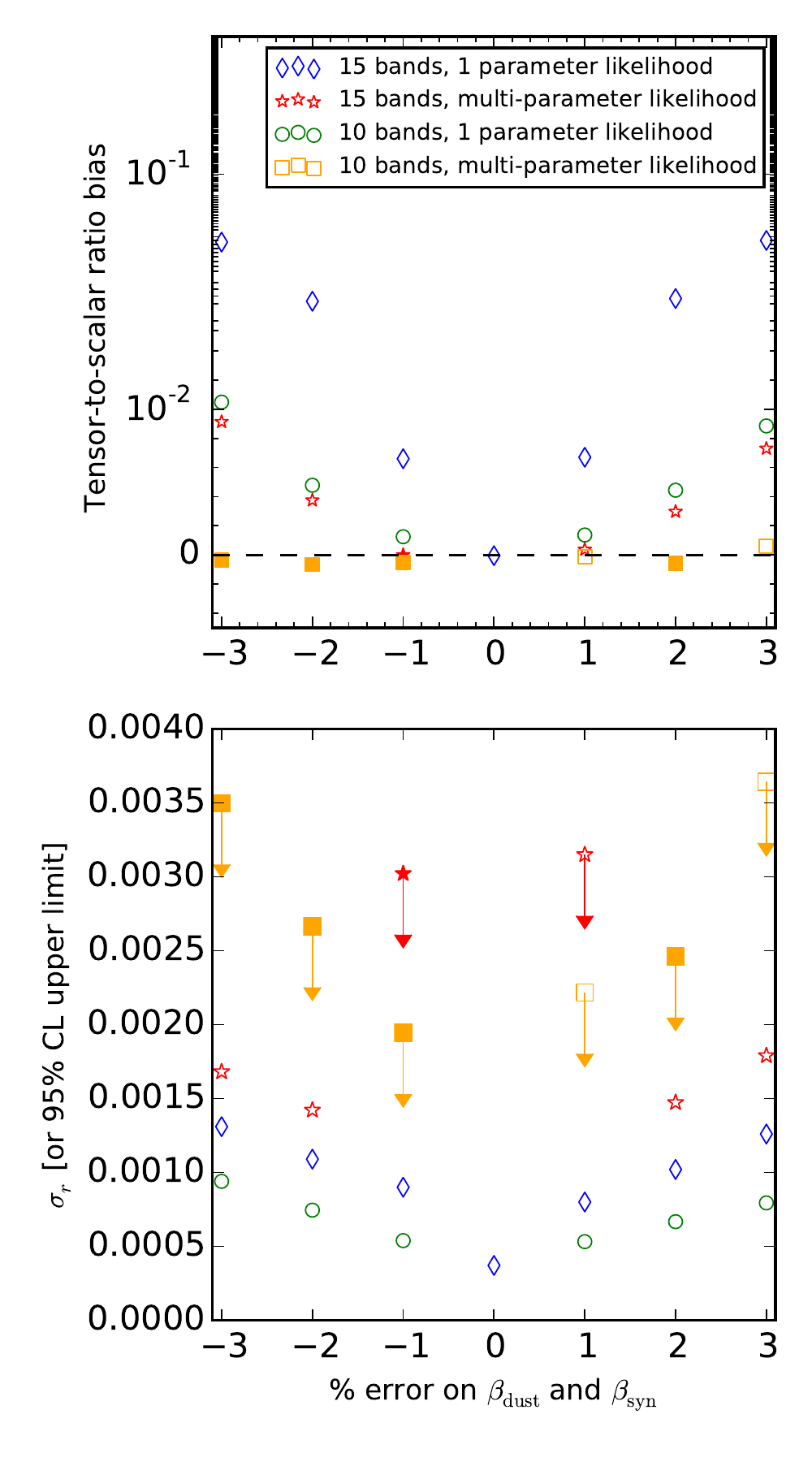}
    \caption{Same as Fig.~\ref{fig:bias_r2_constant} for the simulation with constant spectral indices and $r=0.001$. 
If the measured $r$ value is less than 2\,$\sigma_r$ away from $r=0$, instead we plot the 95\% upper limit, with an arrow down symbol. \label{fig:bias_r3_constant}
}
\end{figure}

\subsection{Sky model with constant spectral indices (Simple model)} \label{sec:constant_constant} We use the model described in Section~\ref{sec:simulations}, where the foregrounds have spatially constant spectral indices ($\beta_{\rm dust}=1.53$ and $\beta_{\rm syn}=3.10$) and we run the component separation assuming fixed errors on these spectral indices. These are $\pm1$, 2 and 3\% errors on both $\beta_{\rm dust}$ and $\beta_{\rm syn}$. As a reference, we also examined the case of perfect knowledge on the foregrounds, that is, 0\% error on the spectral indices.

We consider two cases for the GLS reconstruction: a linear mixture of all the 15 \core{} frequency bands, and one of only the lowest 10 bands, having $\nu \leq 315$\,GHz. This is motivated by the fact that high-frequency bands are strongly contaminated by thermal dust, so including them in the CMB reconstruction increases the dust residuals for a given error on the dust spectral index. The drawback is an increase in the noise level, that needs to be weighted against the reduction in the foreground residuals. In any case, it is worth pointing out that the whole frequency range should be used in order to  estimate the spectral indices before the GLS reconstruction, as this strategy in general achieves the smallest errors on $\beta_{\rm dust}$ and $\beta_{\rm syn}$. 

For all the assumed error cases, we calculate both the likelihood of equation~(\ref{eq:theory_cl}), where the only parameter is $r$, and the multi-parameter likelihood of equation~(\ref{eq:theory_cl_2}), where we include either one or both of the foreground parameters $A_{\rm dust}$ and $A_{\rm syn}$, depending on what is achieving the lowest reduced $\chi^2$ value.



As an example of the multi-parameter likelihood method, we show in Fig.~\ref{fig:ps_constant} the $BB$ power spectrum for the case with $r=0.01$, $+2$\% error on both spectral indices and using only the frequency bands $\leq 315$\,GHz. The reconstructed CMB (shown as the black circles)  contains extra power because of the foreground residuals. However, we are able to model the foreground residuals, shown as diamonds for the thermal dust and as stars for the synchrotron. 
The 3-parameter model yields and unbiased value of $r=0.0088 \pm 0.0020$ despite the large foreground residuals present. The 1-parameter model gave the highly biased result of $r=0.0153 \pm 0.001$ for the same case.

In Fig.~\ref{fig:likelihoods_constant} we show some example tensor-to-scalar ratio likelihoods for the simulations with $r=0.01$ (left) and $r=0.001$ (right). The grey curves show the perfect knowledge component separation using all the 15 frequency bands, which always yields an unbiased result ($r=0.0099 \pm 0.0009$ and $r=0.00095\pm0.00037$).  All the other curves assume a $+3$\% error on both spectral indices. The blue curves correspond to a CMB reconstruction using 15 frequency bands, and a 1-parameter (dashed) and multi-parameter (solid) likelihood. The red curves are the same for a CMB reconstruction using only the first 10 frequency bands.   

For both the $r=0.01$ and $r=0.001$ cases, the multi-parameter likelihood on the 10 frequency bands case allow removing the large bias corresponding to the $+3$\% spectral index error. However, there is a degradation in the measured error $\sigma_r$. For the $r=0.001$ case, this does not allow a detection over 2\,$\sigma$ any more, but only corresponds to an upper limit.

We show the summary of all the results for the simulated observation with $r=0.01$ in top half of Table~\ref{table:constant} and in Fig.~\ref{fig:bias_r2_constant}. In the top panel, we show the measured bias (estimated minus true $r$); in the bottom panel, we show the width of the likelihood, $\sigma_r$. As expected, when we assume perfect knowledge of the foregrounds, the result is unbiased. However, when we introduce some error in the component separation, the likelihood is biased towards higher values of $r$. If we adopt the multi-parameter likelihood instead of the 1-parameter one, the bias is either reduced or removed. 

Limiting the frequency bands used in the CMB solution to $\nu \leq 315$\,GHz is also effective in reducing the bias. In fact, a large fraction of the foreground residuals is introduced by the high frequency bands that are strongly dominated by thermal dust. By comparing the red stars (multi-parameter likelihood, full frequency range) to the green circles (1-parameter likelihood, limited frequency range), we see that they give similar biases for the same error on the spectral parameters. However, the green circles have smaller $\sigma_r$ values than the red stars, this showing that, in this case, it is preferable to limit the bands used in the component separation than to introduce a multi-parameter likelihood. Even so, in some cases, when the bias is large (e.g. $\pm 3$\% spectral index error), both approaches must be used at the same time (shown by the yellow triangles). 

The summary of all the cases for the simulations with $r=0.001$ is reported in the bottom half of Table \ref{table:constant} and shown in Fig.~\ref{fig:bias_r3_constant}. It follows the same scheme from Fig.~\ref{fig:bias_r2_constant}, for the same assumed component separation error cases. Getting an unbiased result is much more difficult in this case, due to the small value of $r$. In particular, for errors in the spectral indices larger than $\pm 1$\%, we always need both the multi-parameter likelihood  and the limited frequency range. We note that, for the perfect knowledge case, the value of $\sigma_r=3.7\times10^{-4}$ is only just below the value allowing a 2\,$\sigma$ detection. Therefore, the multi-parameter likelihood increases $\sigma_r$ and only allows for a 95\% upper limit.


\begin{table*}
\footnotesize
\centering
    \begin{tabular}{l l l l l l l l l l}
    \hline
    \textbf{$r$ value}	&	\textbf{$\Delta \beta_{\rm dust}$,$\Delta \beta_{\rm syn}$} & \multicolumn{4}{c}{\textbf{Using all 15 bands}} & \multicolumn{4}{c}{\textbf{Using 10 bands $\nu \leq 315$\,GHz}} \\
    \hline
    							&									&\multicolumn{2}{c}{1-parameter}&\multicolumn{2}{c}{multi-parameter}&\multicolumn{2}{c}{1-parameter}&\multicolumn{2}{c}{multi-parameter}\\
    							&									&bias&$\sigma_r$ [$10^{-4}$]&bias&$\sigma_r$[$10^{-4}$]&bias&$\sigma_r$[$10^{-4}$]&bias&$\sigma_r$[$10^{-4}$]\\
    \hline
    \multirow{7}{*}{$r=0.01$}	& 0\%	& $1.2\times10^{-4}$(0.1) 	& $8.8$ & -- & -- & -- & -- & -- & -- \\
    							& +1\%	& $6.5\times10^{-3}$(7.0)	& $9.2$ & $1.4\times10^{-3}$(1.1) 	& $12$ 	& $1.3\times10^{-3}$(1.3)& $9.8$ & $-1.6\times10^{-3}(-0.9)$	& $17$ \\
    							& -1\%	& $7.2\times10^{-3}$(6.6)	& $11$	& $5.0\times10^{-5}$(0.0)	& $18$ 	& $1.1\times10^{-3}$(1.0)& $11$ & $-1.5\times10^{-3}(-0.9)$	& $16$ \\
                                & +2\%	& $2.2\times10^{-2}$(20)	& $11$ 	& $5.2\times10^{-3}$(3.6)	& $15$ 	& $5.3\times10^{-3}$(5.0)& $10$ & $-1.1\times10^{-3}(-0.5)$ 	& $21$ \\
                                & -2\%	& $2.4\times10^{-2}$(18)	& $13$ 	& $4.7\times10^{-3}$(2.5) 	& $19$	& $4.4\times10^{-3}$(3.7)	& $12$ & $-8.8\times10^{-4}(-0.5)$	& $19$ \\
                                & +3\%	& $4.2\times10^{-2}$(32)	& $13$ 	& $1.1\times10^{-2}$(6.2) 	& $17$ 	& $1.1\times10^{-2}$(9.3)& $12$ & $1.0\times10^{-3}$(0.5)	& $19$ \\
                                & -3\%	& $4.7\times10^{-2}$(31) 	& $15$ 	& $9.9\times10^{-3}$(4.6) 	& $22$ 	& $9.4\times10^{-3}$(7.0)& $13$ & $-1.6\times10^{-4}(-0.1)$	& $21$ \\
                                
	\hline
    \multirow{7}{*}{$r=0.001$}	& 0\%	& $-5.0\times10^{-5}(-0.1)$	& 3.7 	& -- 			& -- 					& -- & -- & -- & -- \\
    							& +1\%	& $6.7\times10^{-3}$(8.3) 	& 8.0 	& $3.7\times10^{-4}$ 		& $<31.5^{\dagger}$ & $1.4\times10^{-3}$(2.6) 	& 5.3 & $-1.1\times10^{-4}$	& $<22.2^{\dagger}$ \\
    							& -1\%	& $6.6\times10^{-3}$(7.3)	& 9.0 	& $0.0$ 					& $<30.2^{\dagger}$ & $1.3\times10^{-3}$(2.3)	& 5.4 & $-5.3\times10^{-4}$	& $<19.4^{\dagger}$ \\
                                & +2\%	& $2.3\times10^{-2}$(23)	& 10 	& $3.0\times10^{-3}$(2.0) 	& 15 				& $4.4\times10^{-3}$(6.6)	& 6.7 & $-5.9\times10^{-4}$	& $<24.6^{\dagger}$ \\
                                & -2\%	& $2.3\times10^{-2}$(21) 	& 11 	& $3.8\times10^{-3}$(2.6)	& 14 				& $4.8\times10^{-3}$(6.4)	& 7.5 & $-6.5\times10^{-4}$	& $<26.7^{\dagger}$ \\
                                & +3\%	& $4.6\times10^{-2}$(37)	& 13 	& $7.3\times10^{-3}$(4.1)  	& 18 				& $8.9\times10^{-3}$(11)	& 7.9 &	$5.8\times10^{-4}$ 	& $<36.4^{\dagger}$ \\
                                & -3\%	& $4.5\times10^{-2}$(34)	& 13	& $9.1\times10^{-3}$(5.4)	& 17 				& $1.0\times10^{-2}$(11)	& 9.4 & $-3.6\times10^{-4}$	& $<35.0^{\dagger}$ \\                        
	\hline
    \end{tabular}
    \caption{Measured tensor-to-scalar ratio biases and $\sigma_r$ values for runs on the simulation with spatially constant spectral indices. The component separation is modelled using the true spectral indices with a small error. In the $r$ bias columns, the bias expressed as number of $\sigma_r$ is shown in parenthesis. The values with $^{\dagger}$ are 95\% upper limits. \label{table:constant}}
\end{table*}


\subsection{Sky model with spatially variable spectral indices (Complex model)} Now, we consider a more realistic model of the sky, where the spectral indices of the foreground components are spatially variable, as explained in Section~\ref{sec:simulations}. We simulate the measurement of the tensor-to-scalar ratio for two levels of component separation modelling complexity, as detailed in the two following sub-sections. 



\begin{figure}
	\centering
    \includegraphics[width=1\columnwidth]{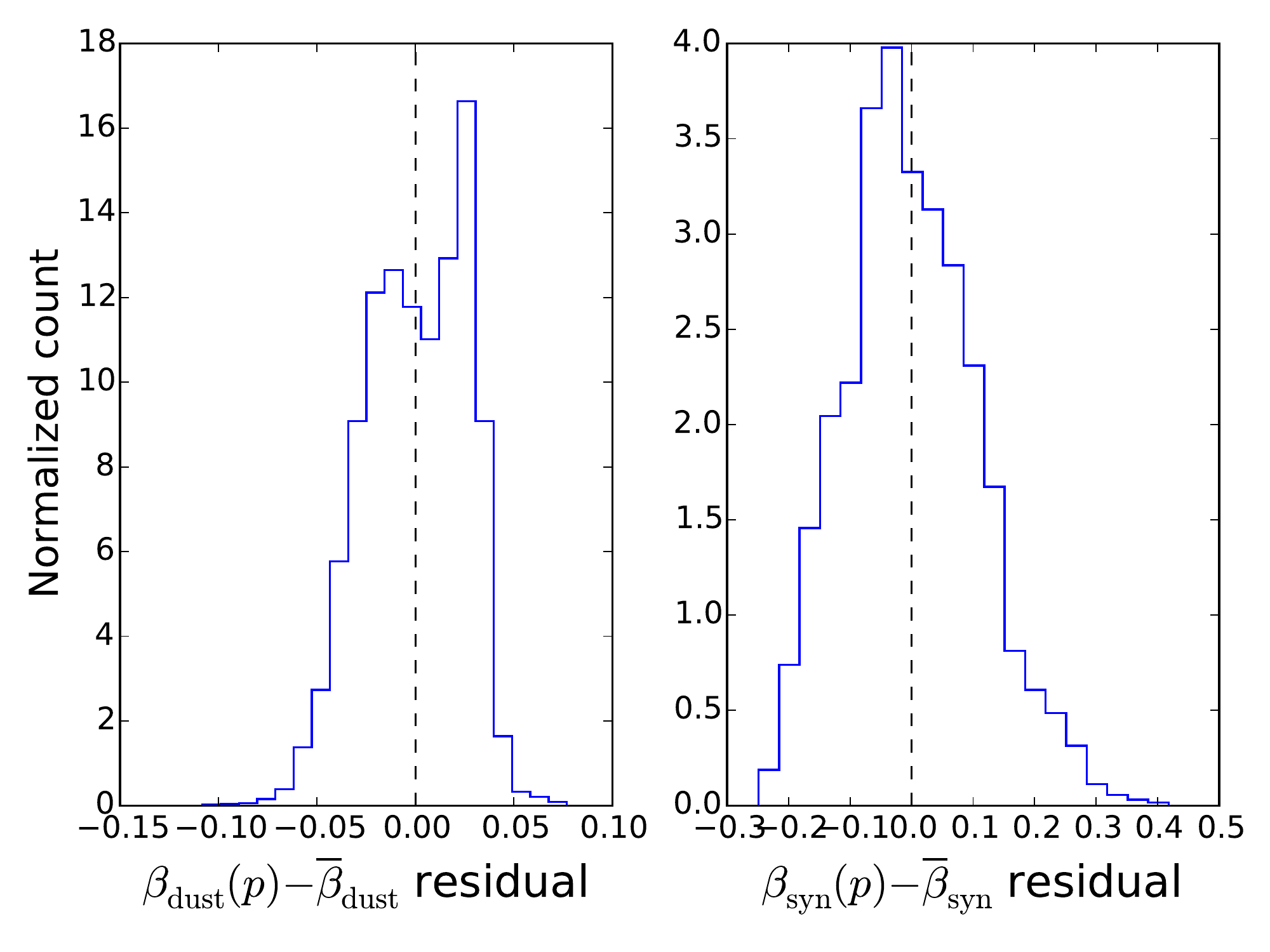}
    \caption{
    Histograms of the dust (left) and synchrotron (right) spectral index residuals  (true -- average spectral index) outside the Galactic mask. The true spectral indices are the ones in the sky model from \citet{hervias_2016}. The true average $\beta_{\rm dust}$ is calculated on a $N_{\rm side}=2048$ map, while the true average $\beta_{\rm syn}$ is calculated on a $N_{\rm side}=512$ map. The histograms are normalized so that they integrate to 1. The standard deviations of the distribution of spectral index residuals are equivalent to a 1.7\% error for thermal dust and to a 3.5\% error for synchrotron. \label{fig:histograms}}
\end{figure}

\subsubsection{Modelling the component separation with spatially constant $\beta_{\rm dust}$ and $\beta_{\rm syn}$} \label{sec:variable_constant} The first approach we adopt is to model the spatially variable spectral indices as a constant value across the sky. We set this value to the average of the true $\beta_{\rm dust}$ and $\beta_{\rm syn}$ maps outside the Galactic mask of Fig.~\ref{fig:base_mask}. As such, these values ($\bar{\beta}_{\rm dust}=1.53$  and $\bar{\beta}_{\rm syn}=2.89$) better represent most of the pixels used for the analysis. The histograms of the spectral index residuals, defined as the difference between the true indices (at a given pixel) and the true average value,
for all the pixels in the sky outside the mask are shown in Fig.~\ref{fig:histograms}. The standard deviation of these residuals are 0.0253 for $\beta_{\rm dust}$ and 0.1074 for $\beta_{\rm syn}$, which corresponds to a 1.7\,\% and 3.5\,\% error, respectively. These errors are qualitatively similar to the ones we considered in Section~\ref{sec:constant_constant}.

\begin{figure}
	\centering
    \includegraphics[width=1.0\columnwidth]{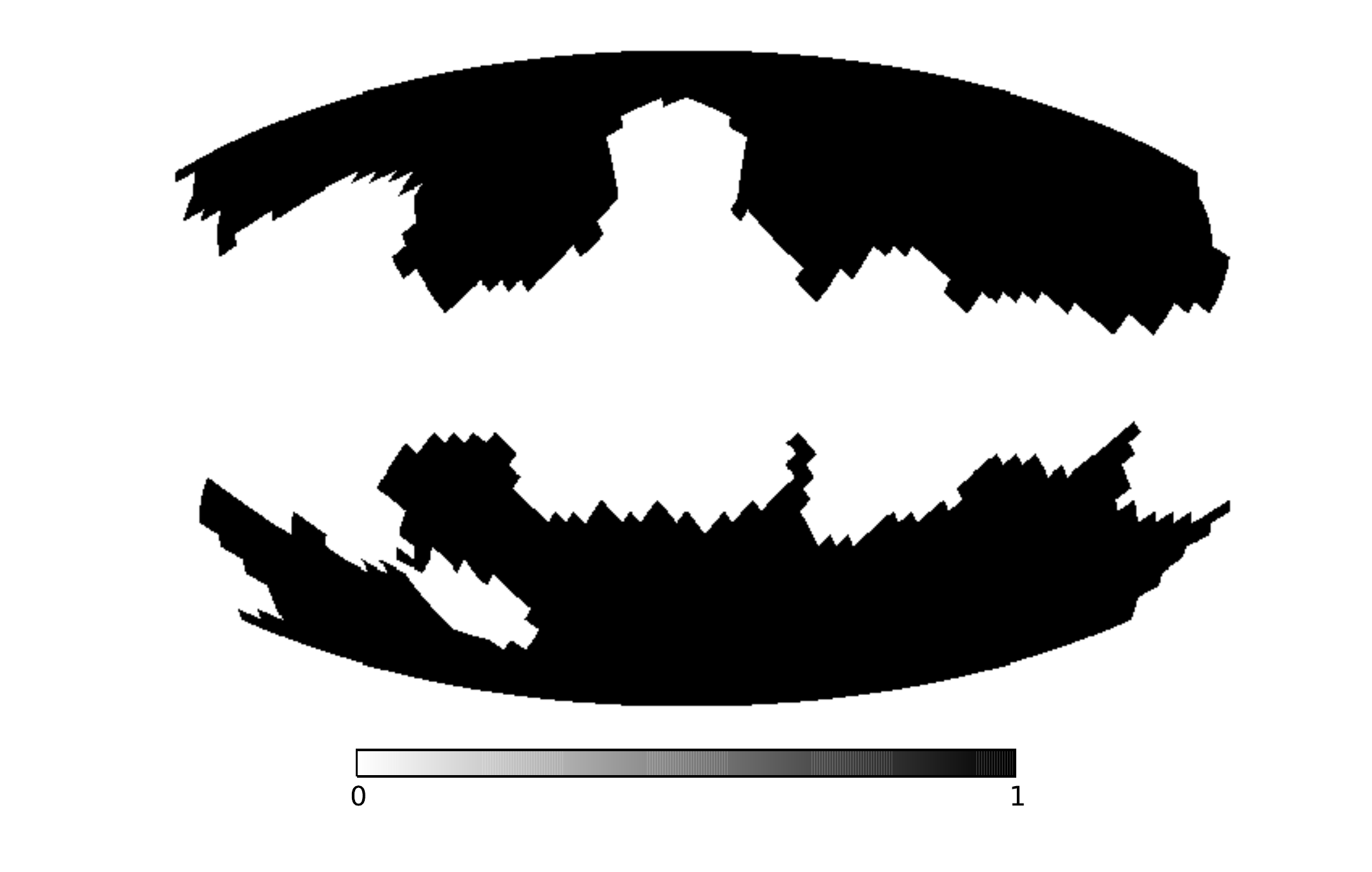}
    \includegraphics[width=1.0\columnwidth]{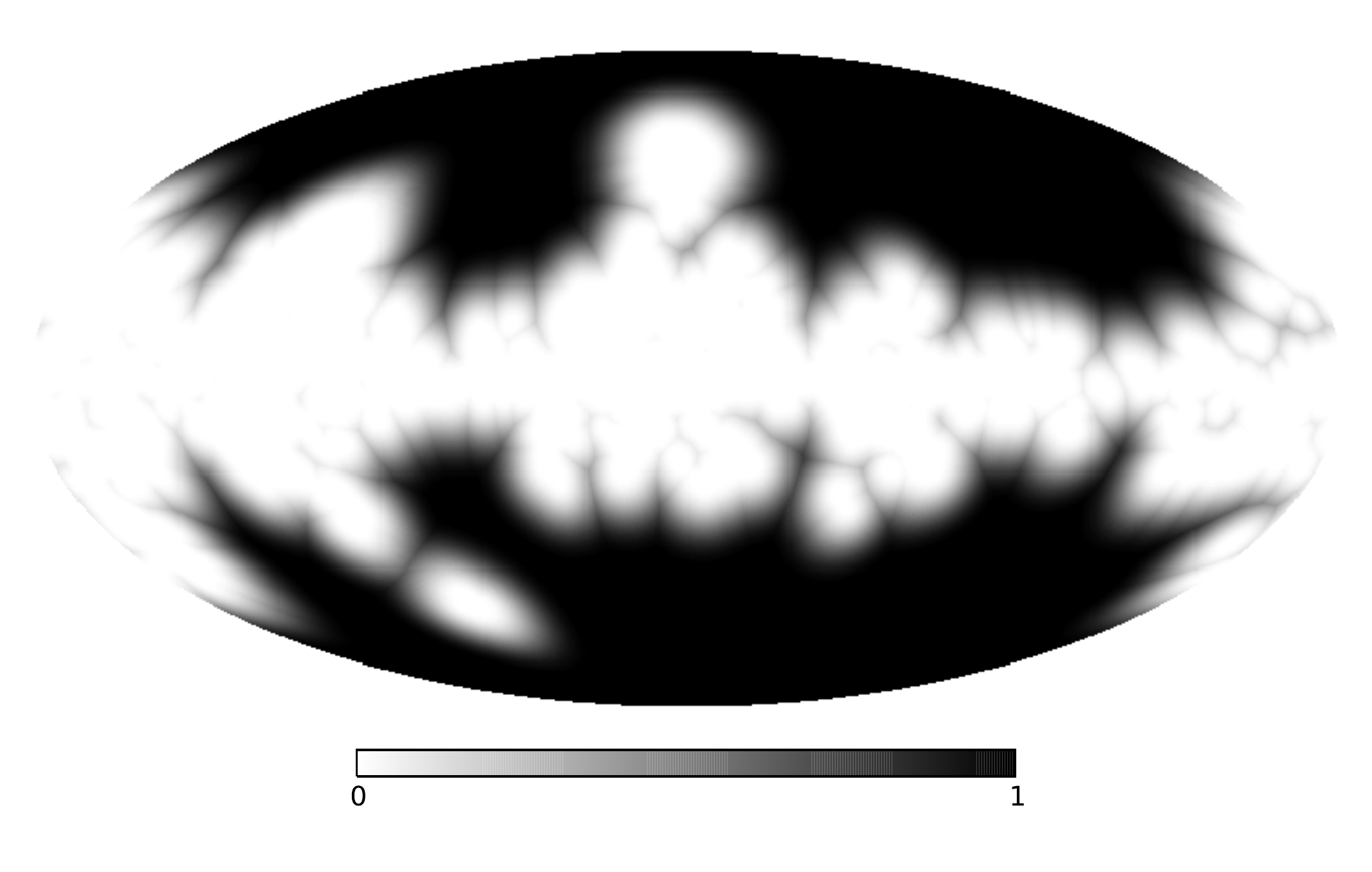}
    \caption{Same as Fig.~\ref{fig:base_mask}, but now showing the optimized Galactic mask for the runs described in Section~\ref{sec:variable_constant} and labelled as \emph{best} in the top half of Table~\ref{table:results_variable}. The mask has $f_{\rm sky}=0.48$. \label{fig:adhoc_mask}}
\end{figure}

We start by considering the 1-parameter and multi-parameter likelihood on the CMB reconstructed using the first 10 frequency bands. This was the best-performing case in the previous (constant spectral index) exercise. The results are reported in the top half of Table~\ref{table:results_variable} as the \emph{base} case. As we can see, this case is not good enough any more, because the bias on $r$ is still significant. We therefore proceeded to optimize the analysis to reduce the bias on $r$, and obtained the results quoted in Table~\ref{table:results_variable} as the \emph{best} case. The modifications we introduced are:

\begin{enumerate}

\item \textbf{Secondary dust component}: we use the idea presented in \cite{stolyarov_2005}, which is that a component with spatially variable spectral dependence can be modelled as a series of components with constant spectral dependence, each one corresponding to a term in a Taylor expansion. In this way, the first-order thermal dust is the usual grey-body spectral law, and we add a second-order thermal dust component, whose spectrum is the derivative of that spectral law with respect to $\beta_{\rm dust}$. We also explored the possibility to add a second synchrotron component to account for the variability of the synchrotron spectral index, but this achieved no significant improvement in our case.
 
\item \textbf{Optimization of the Galactic mask}: we produced a new mask specifically optimized for $B$-modes and tailored to the component separation approach we used. Specifically, we produced an estimate of foreground errors on the CMB $B$ map with a Monte Carlo (MC) approach, and then excluded all pixels for which this map was over some threshold. To derive the error $B$ map, we repeated the GLS CMB reconstruction 100 times by varying randomly the assumed spectral indices with Gaussian distributions (having $\sigma_{\beta_{\rm dust}} = \sigma_{\beta_{\rm syn}}=0.1$ and $\sigma_{T_{\rm dust}}=1$\,K). We transformed each MC output CMB from $Q/U$ to $E/B$ and finally computed the standard deviation of the 100 MC $B$ maps for each pixel. The optimized Galactic mask used in the \emph{best} case analysis is shown in Fig.~\ref{fig:adhoc_mask}. This mask has $f_{\rm sky}=0.48$, which is very similar to that of the mask used before.
\end{enumerate}


\begin{figure*}
	\begin{center}
    	\includegraphics[width=0.5\textwidth]{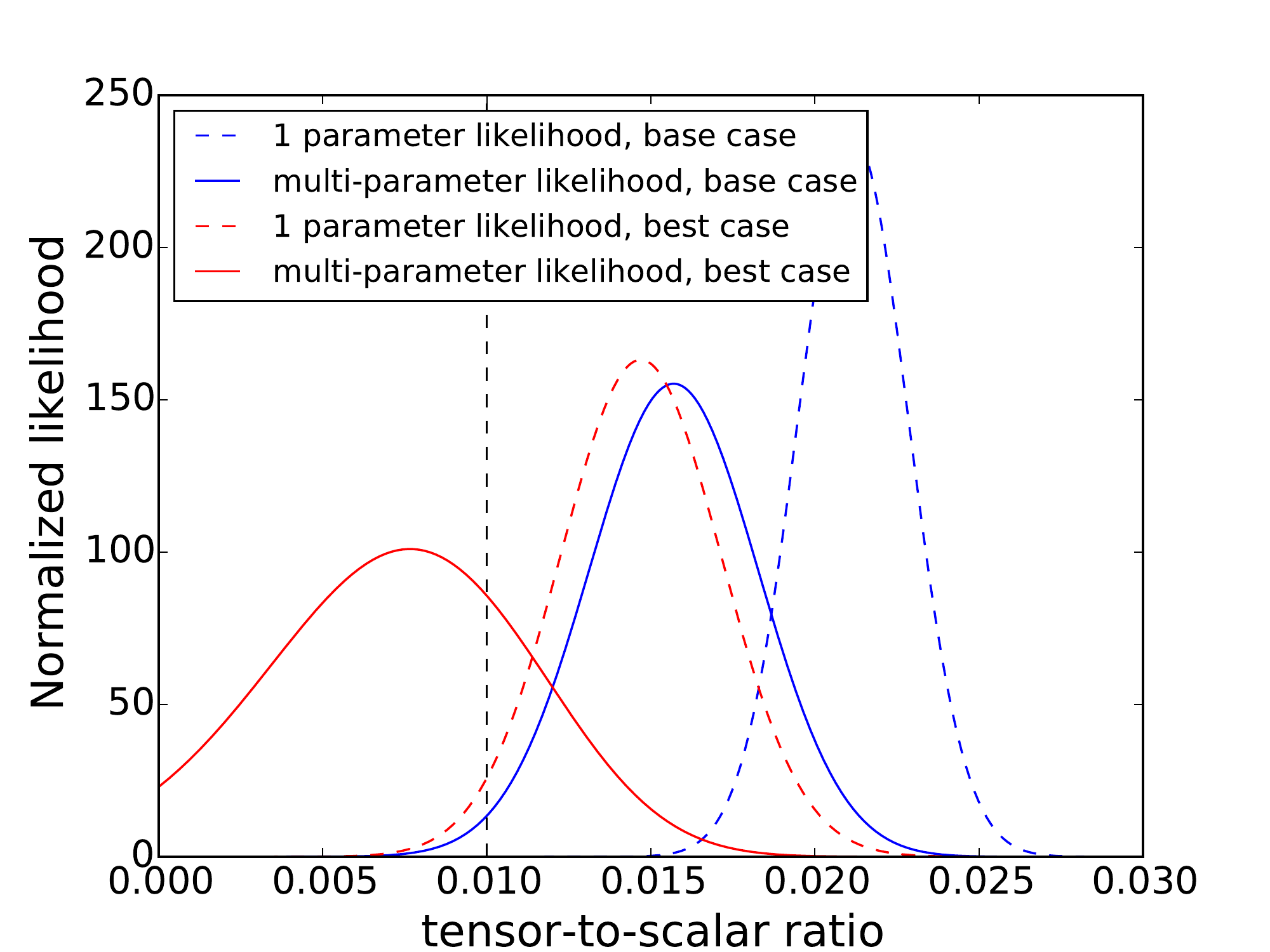}~
        \includegraphics[width=0.5\textwidth]{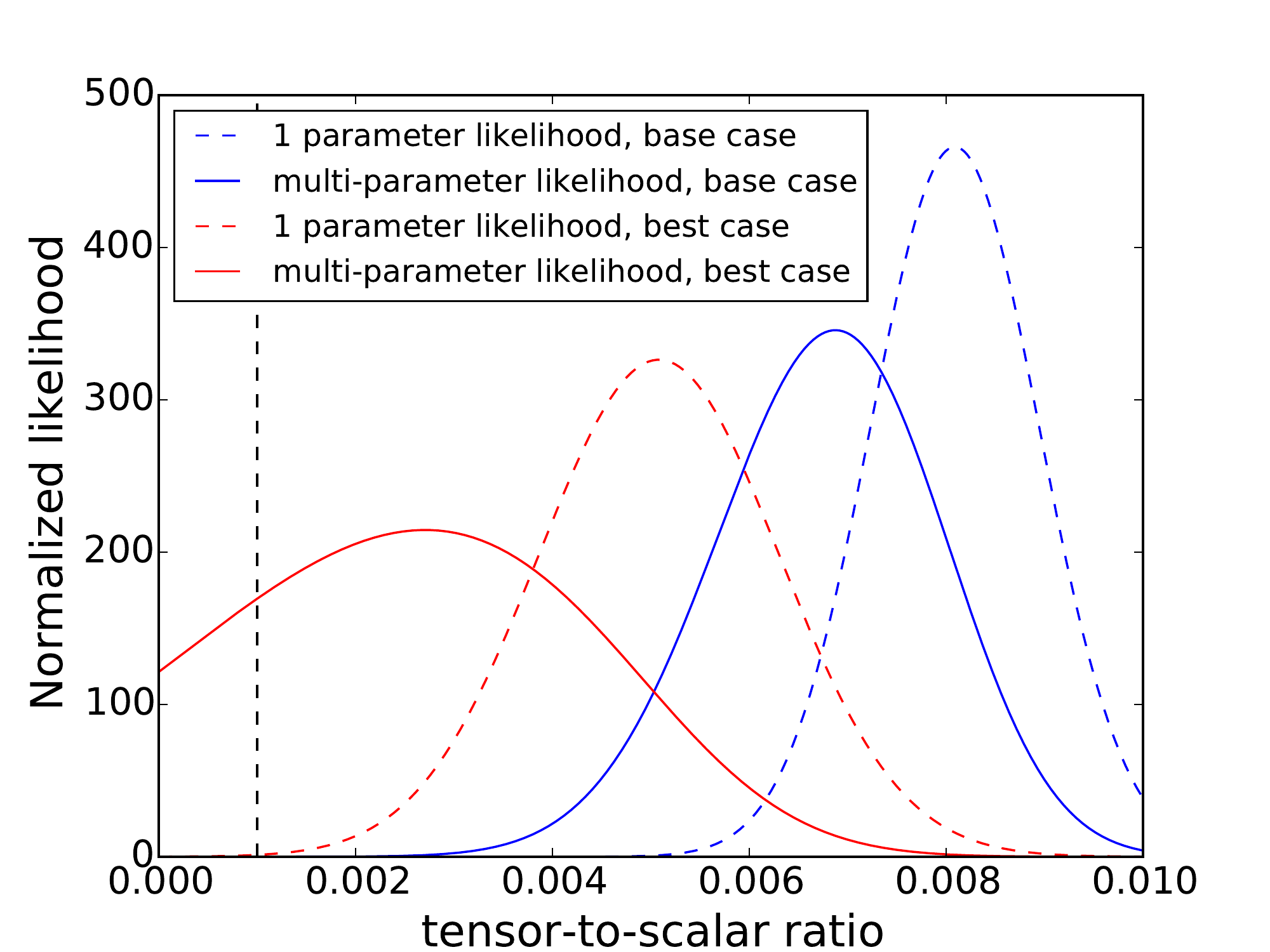}~
	\end{center}
	\caption{Same as Fig.~\ref{fig:likelihoods_constant} for the simulation with spatially varying spectral indices and component separation assuming spatially constant spectral indices, for $r=0.01$
 (left) and $r=0.001$ (right). \label{fig:likelihoods_variable_constant}}
\end{figure*}

The results for this run are reported in the top of Table \ref{table:results_variable} and shown in Fig. ~\ref{fig:likelihoods_variable_constant}. The left panel shows the measured likelihoods for the simulated observations with $r=0.01$. The \emph{base} case measurements (the blue curves) are biased, even when foreground residuals are modelled in the likelihood. However, the improvements we introduced to the analysis allow for an unbiased detection (shown by the solid red curve). This detection is only 2\,$\sigma_r$ away from $r=0$.

Fig.~\ref{fig:likelihoods_variable_constant}, right, shows the same result likelihoods for $r=0.001$. In this case, the results are all biased, even with the improvements in the \emph{best} case (the red curves). This means that, for such a small value of $r$, the systematic error we commit by neglecting the spatial variability of the spectral indices is too big to be compensated.  In this case,  the spatial variability needs to be modelled directly in the component separation, as we do in the next subsection.


\begin{table*}
\centering
    \begin{tabular}{ l l l l l l}
    \hline
    \textbf{$r$ value}		& \textbf{Case}	& \multicolumn{4}{c}{\textbf{Measured values}} \\
    \hline
    			& & \multicolumn{2}{c}{1-parameter}			& \multicolumn{2}{c}{multi-parameter}	\\
     			& & bias  & $\sigma_r$ 	& bias  & $\sigma_r$ \\
    \hline
   	\multirow{2}{*}{$r=0.01$} 	& Base 	& $1.1\times10^{-2}(6.6)$ 	& $1.7\times10^{-3}$ & $5.7\times10^{-3}(2.2)$ 		& 0.00257					\\
  								& Best	& $4.7\times10^{-3}(1.9)$ 	& $2.5\times10^{-3}$ & $-2.4\times10^{-3}(-0.6)$	& 0.00390					\\
    \hline
    \multirow{2}{*}{$r=0.001$}	& Base	& $7.1\times10^{-3}(8.2)$	& $8.6\times10^{-4}$ & $5.9\times10^{-3}(5.1)$		& 0.00115					\\
    							& Best	& $4.1\times10^{-3}(3.3)$ 	& $1.2\times10^{-3}$ & $1.7\times10^{-3}$			& $<0.00566^{\dagger}$ 		\\
    \hline \noalign{\smallskip}
    
    \textbf{$r$ value}			& \textbf{$\Delta \beta_{\rm dust}$,$\Delta \beta_{\rm syn}$ global error} & 	\\
    \hline
    \multirow{5}{*}{$r=0.001$} 	& 0,0	\%	& $1.0\times10^{-5}(0.0)$	& $4.5\times10^{-4}$ 	& -- 					& -- 		\\
    							& 1,1 \%	& $4.6\times10^{-4}(0.9)$	& $5.2\times10^{-4}$ 	& $-1.2\times10^{-4}$ 	& $<1.9\times10^{-3}$ $^{\dagger}$ \\
    							& 0.5,0.5\%	& $2.1\times10^{-4}(0.4)$	& $4.9\times10^{-4}$ 	& -- 					& --		\\
    							& 1,0.5	\%	& $3.1\times10^{-4}(0.6)$	& $4.9\times10^{-4}$ 	& -- 					& --		\\
    							& 0.5,1	\%	& $2.3\times10^{-4}(0.5)$	& $4.9\times10^{-4}$ 	& --	 				& -- 		\\    
    \hline
    \end{tabular}
    \caption{Measured tensor-to-scalar ratio biases and $\sigma_r$ values for runs on the simulation with variable spectral indices. On the top half, we show the results for modelling with constant spectral indices in the component separation. In the bottom half, we show the results for modelling with the true spatially variable spectral indices with a small level of error in the component separation. In the $r$ bias columns, the bias expressed as number of $\sigma_r$ is shown in parenthesis. The values with $^{\dagger}$ are 95\% upper limits. \label{table:results_variable}}
\end{table*}


\subsubsection{Modelling the component separation with spatially variable spectral indices} \label{sec:variable_variable} For the model with $r=0.001$, which did not give an unbiased result in the previous subsection, we model the spatial variability of the spectral indices directly into the component separation. Indeed, most component separation methods are able to perform a local estimation of the foreground spectral properties, either pixel-by-pixel (e.g. \textit{commander}, \citealt{eriksen_2008}, \textit{MIRAMARE}, \citealt{stompor2009}), on sky patches (e.g. \textit{CCA}, \citealt{ricciardi2010}) or by means of other kind of spatial localization \citep[e.g., \text{NILC}][]{delabrouille_2009_nilc,basak2013}. As a drawback, estimation errors might be larger on a local estimation than on a global one, especially where foregrounds are weaker.

That is, in lines of sights where the (polarized) intensity is stronger, the error in the determination of spectral properties would in general be smaller, since there is a higher signal-to-noise ratio.

In order to model such error properties in our analysis, we investigate the spatial correlation of errors in $\beta_{\rm dust}$ and $\beta_{\rm syn}$ made with the \textit{commander} algorithm. These simulated observations were produced with the Planck Sky Model \citep[PSM, ][]{PSM2013} for the ``Exploring Cosmic Origins with CORE'' foregrounds paper (CORE collaboration et al. in prep.), and using a Galactic mask with $f_{\rm sky}=0.54$. Although the instrumental specifications and the sky model are slightly different from what we use here, this allows us to derive the basic error properties that are needed for our modelling.

We find that the error on the synchrotron spectral index, $\Delta \beta_{\rm syn}$, is consistent with a random distribution and it is not significantly correlated with the synchrotron polarized intensity. This is because of the frequency coverage of \textit{CORE}, which does not include many synchrotron dominated channels, which means that the synchrotron estimation error is essentially noise-dominated. We then modelled the synchrotron spectral index as having a Gaussian distribution with standard deviation:
\begin{equation}
	\sigma_{\Delta \beta_{\rm syn}} = \epsilon \bar{\beta}_{\rm syn}/100 \label{eq:betasyn_error},
\end{equation}
where $\epsilon$ is the error percentage we assume in our analysis and $\bar{\beta}_{\rm syn}=2.89$ is the average synchrotron spectral index outside the Galactic mask. 

We find that the thermal dust error, $\Delta \beta_{\rm dust}$, is clearly anti-correlated with the polarized dust intensity. To model this property, we binned the pixels outside the Galactic mask into ranges of polarized intensity $P$ having roughly the same number of data points and fitted Gaussian density functions to the error distribution in each bin. The standard deviation as a function of $P$ is well modelled by a power-law:
\begin{equation}
	\sigma_{\Delta \beta_{\rm dust}}(P) = \epsilon A_{1\%} (P/\mu{\rm K}_{\rm 353 GHz})^{-b} \text{,} \label{eq:power_law}
\end{equation}
where $A_{1\%}=0.019 \pm 0.003$ is the normalization corresponding to $\sigma_{\Delta \beta_{\rm dust}}=\bar{\beta}_{\rm dust}/100$ outside the Galactic mask, $b=0.019 \pm 0.003$ is the slope of the anti-correlation of $\sigma_{\Delta \beta_{\rm dust}}$ with $P$, and $\epsilon$ is the assumed error percentage.

To simulate the estimation of spatially variable spectral indices on our study, we start from the true input spectral indices maps, and add random error maps consistent with the error characterization of eqns. (\ref{eq:betasyn_error}) and (\ref{eq:power_law}). We assume error levels of $1\,\%$ ($\epsilon=1$) and $0.5\,\%$ ($\epsilon=0.5$).

We generated 100 realizations of both $\Delta \beta_{\rm syn}$  and $\Delta \beta_{\rm dust}$ at $N_{\rm side}=16$. We run our component separation pipeline on each of the 100 simulation sets, having a different CMB, noise and random spectral index error realization, on the $\nu \leq 315$\,GHz limited frequency range. We used the $N_{\rm side}=16$ spectral index maps directly for the low-resolution pipeline, and we upgraded them to $N_{\rm side}=512$ for the high-resolution pipeline.

\begin{figure}
	\centering
    \includegraphics[width=1\columnwidth]{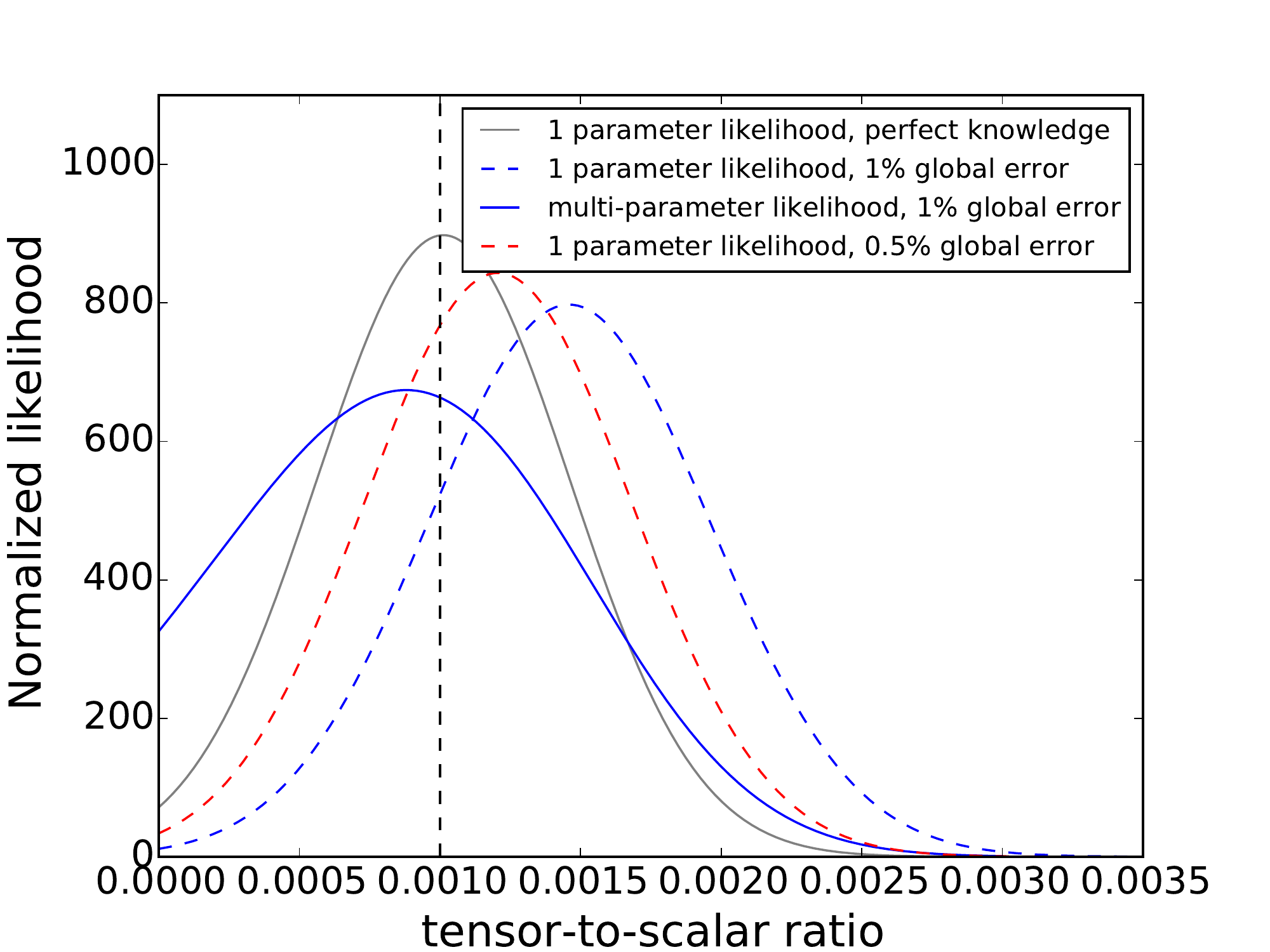}
    \caption{Tensor-to-scalar ratio likelihoods for a complex model (with spatially variable spectral indices) and modelling the component separation as spatially variable spectral indices. All the runs are made with the optimized mask shown in Fig.~\ref{fig:adhoc_mask} and limiting the frequency coverage to $\nu \leq 315$\,GHz. The case with 1\% global error has a 1\% error (standard deviation of error in pixels outside the Galactic mask) modelled with a spatially uniform random Gaussian error for $\Delta \beta_{\rm syn}$ and  with a spatially correlated (following equation~\ref{eq:power_law}) random Gaussian error for  $\Delta \beta_{\rm syn}$. The case with 0.5\% global error is analogous. \label{fig:likelihoods_var_var}}
\end{figure}



The likelihoods for the reconstructed power spectrum averaged over the 100 realizations for both the 1\% and 0.5\% global error cases, along with the case assuming perfect knowledge on the spectral indices, are shown in Fig.~\ref{fig:likelihoods_var_var} and reported in the bottom part of Table \ref{table:results_variable}. As usual, assuming a perfect knowledge (spectral indices errors equal to 0) yields an unbiased result, $r=0.00101 \pm 0.00045$. With a 1\% global error on both spectral indices, we measure a bias on $r$ of $4.6\times10^{-4}$, and a $\sigma_r=5.2\times10^{-4}$ for the 1-parameter likelihood. The multi-parameter likelihood yields a very small bias of $-1.2\times10^{-4}$, but with a 95\% upper limit of $0.00189$.


With 0.5\% global error, we still measure a small bias, of $2.1\times10^{-4}$ and an error of $4.9\times10^{-4}$ for the 1-parameter likelihood. Fitting with the foregrounds nuisance parameters is not well motivated, since the marginalized likelihoods for $A_{\rm dust}$ and $A_{\rm syn}$ are consistent with 0.


We have verified that the residual systematic error on $r$ is due in a greater proportion by thermal dust than to synchrotron residual contamination. In fact, the 1-parameter likelihood yields $r=(13.1\pm4.9)\times10^{-4}$, and $r=(12.3\pm 4.9)\times10^{-4}$, respectively, if we consider a 0.5\% global error only on $\beta_{\rm syn}$ and $\beta_{\rm dust}$ and we leave the other foreground at 1\% global error.

\section{Conclusions} \label{sec:conclusions} We have performed an analysis on the tensor-to-scalar ratio bias produced by the mis-modelling of foreground spectral parameters, taking into account a realistic model of the sky and a full data analysis pipeline.
We have considered two sky models: a very simple one, where the foregrounds (synchrotron and dust) have constant frequency spectra across the sky, and a more complex one, where the spectral dependence is spatially-varying. We modelled component separation strategies and likelihood estimations of increasing complexity.  The main results of our analysis can be summarized as follows.

For $r=0.01$, we obtain an unbiased estimation of $r$ for all simulations considered. The requirements on the accuracy of foregrounds modelling for component separation purposes are not too stringent (for example modelling spatially-varying foreground spectral indices as spatially-constant still gives a successful measurement).

Depending on the error level on the synchrotron and thermal dust spectral indices (from 1\% to 3\%), the best results may require exploiting a limited set of ``cleaner'' frequency maps to reconstruct the CMB. The use of all channels is anyway recommended to obtain an accurate estimation of the foreground spectral indices. We achieve significant improvements by explicitly modelling the synchrotron and dust foreground residuals in the likelihood, and marginalizing over foreground amplitude nuisance parameters. Furthermore, an important role is played by the Galactic mask, that needs to be optimized for the component separation method used and for $B$-mode detection.

For $r=0.001$ and a simple sky model, using a suitable mask, modelling foreground residuals in the likelihood and limiting the frequency range for CMB reconstruction always yields an unbiased $r$ value. The error on $r$ often does not allow a detection but just an upper limit; this result is however conservative because the Gaussian likelihood we adopted is not optimal in the low-multipole regime. 

When increasing the complexity of the sky, large modelling errors, such as approximating a spatially-varying spectral index with a constant,  are not allowed in this case, as they give raise to biases on $r$ that are too large to be corrected for (at the likelihood level). Modelling the spatial variability of the spectral indices at the component separation level is required. For a global error of 0.5--1\% in $\beta_{\rm dust}$ and $\beta_{\rm syn}$, we obtain an unbiased detection/upper limit on $r$. We show that the foreground residuals biasing the measurement of $r=0.001$ are due in greater proportion to thermal dust emission than synchrotron emission, due to the particular frequency coverage of \core{}.

Such level of accuracy in the determination of foreground spectral parameters is very challenging, and motivates further research on component separation and foreground characterization. However, our analysis does not take into account polarization ancillary data that will become available, such as C-BASS \citep{cbass_paper}. Our method could also be used to further optimize the instrumental specifications of future CMB B-modes experiments.

\section*{Acknowledgements} We thank M. Remazeilles for his kind help with the \textit{commander} code. We thank the CORE collaboration for allowing us to use data to calculate spectral parameter errors. CHC acknowledges the funding from Becas Chile/CONICYT. AB and MLB acknowledge support from the European Research Council under the EC FP7 grant number 280127. MLB also acknowledges support from an STFC Advanced/Halliday fellowship. We gratefully acknowledge the anonymous referee for useful suggestions that led to the improvement of this paper.


\bibliographystyle{mnras}
\bibliography{biblio} 







\bsp	
\label{lastpage}
\end{document}